\documentclass[twocolumn]{aastex63}

\usepackage{epsfig}
\usepackage{xcolor}
\usepackage{amssymb}
\usepackage{pifont}
\newcommand{\cmark}{\ding{51}}
\newcommand{\xmark}{\ding{55}}
\usepackage{romannum}
\usepackage{lipsum}

\received{}
\revised{}
\accepted{}
\submitjournal{ApJ}

\shorttitle{SMBHB Detection and Parameter Estimation with PTAs}
\shortauthors{T.~Liu and S.~J.~Vigeland}

\begin{document}

\title{
Multi-messenger Approaches to Supermassive Black Hole Binary Detection and Parameter Estimation: Implications for Nanohertz Gravitational Wave Searches with Pulsar Timing Arrays
}

\correspondingauthor{Tingting Liu}
\email{tingtliu@uwm.edu}

\author{Tingting Liu}
\affiliation{Center for Gravitation, Cosmology and Astrophysics, Department of Physics, University of Wisconsin-Milwaukee, Milwaukee, WI 53211, USA}

\author{Sarah J.~Vigeland}
\affiliation{Center for Gravitation, Cosmology and Astrophysics, Department of Physics, University of Wisconsin-Milwaukee, Milwaukee, WI 53211, USA}


\begin{abstract}
Pulsar timing array (PTA) experiments are becoming increasingly sensitive to gravitational waves (GWs) in the nanohertz frequency range, where the main astrophysical sources are supermassive black hole binaries (SMBHBs), which are expected to form following galaxy mergers. Some of these individual SMBHBs may power active galactic nuclei, and thus their binary parameters could be obtained electromagnetically, which makes it possible to apply electromagnetic (EM) information to aid the search for a GW signal in PTA data. In this work, we investigate the effects of such an EM-informed search on binary detection and parameter estimation by performing mock data analyses on simulated PTA datasets. We find that by applying EM priors, the Bayes factor of some injected signals with originally marginal or sub-threshold detectability (i.e., Bayes factor $\sim 1$) can increase by a factor of a few to an order of magnitude, and thus an EM-informed targeted search is able to find hints of a signal when an uninformed search fails to find any. Additionally, by combining EM and GW data, one can achieve an overall improvement in parameter estimation, regardless of the source's sky location or GW frequency. We discuss the implications for the multi-messenger studies of SMBHBs with PTAs.
\end{abstract}


\section{Introduction} \label{sec:intro}

Multi-messenger astrophysics (MMA) began with the observations of neutrinos from the Sun (which dates back to the 1960s) and later from the supernova SN1987A and the 2017 observations of a binary neutron star merger \citep{Abbott2017} ushered in a new era of the field: for the first time, the messengers were light and gravitational waves (GWs). This burgeoning field of GW MMA is enabling new and unique insight into the source's astrophysics that is otherwise inaccessible with traditional observational astronomy, as well as gravity theories, fundamental physics, and cosmology.

The future prospects of MMA with GW and electromagnetic(EM) observations are tremendously promising. The Laser Interferometer Space Antenna (LISA, \citealt{Amaro-Seoane2017}) will observe GWs in the mHz regime ($10^{-4}-10^{-2}$ Hz), whose astrophysical sources include white dwarf binaries, massive black hole binaries (MBHBs, $\gtrsim10^{5}M_{\odot}$), and extreme mass-ratio inspirals (e.g., \citealt{Klein2016,Babak2017}). Pulsar timing array (PTA) experiments (e.g., \citealt{NANOGrav,EPTA,PPTA}) probe the nanohertz regime ($10^{-9}-10^{-6}$ Hz), where the main astrophysical sources are the supermassive black hole binaries (SMBHBs, $\gtrsim10^{9} M_{\odot}$). Unlike most stellar-mass black hole binaries, MBHB and SMBHB systems are expected to emit EM radiation: the galaxy merger process, which is thought to be the formation channel of MBHBs and SMBHBs, can funnel a large amount of gas to the nuclear region \citep{Barnes1992}, thereby enabling accretion onto the BHs and potentially powering active galactic nucleus (AGN) activity.

As these close SMBHB systems in the GW regime (which are typically at approximately milliparsec separations) are largely beyond the reach of direct imaging and interferometric observations, the indirect observational signatures of these binary systems have been of great interest to a large body of theoretical and numerical work. These EM indicators of a close SMBHB can be roughly classified into one or more of the following categories: peculiar spectral features (e.g., \citealt{Gultekin2012, Kocsis2012,Tanaka2012,McKernan2013,Roedig2014,Farris2015,d'Ascoli2018}), periodic variations or flares as the result of an orbiting binary (e.g., \citealt{MacFadyen2008,Noble2012,Shi2012,D'Orazio2013,Farris2014,Gold2014,Tang2018,Bowen2018,D'Orazio2018,Noble2021}), and other associated transient phenomena in an otherwise quiescent system (such as tidal disruption events with modified dynamics, e.g., \citealt{FKLiu2009,Ricarte2016}). Binary population estimates have further made predictions for the occurrence rates of binaries displaying these signatures \citep{Kelley2019,Krolik2019}, which are estimated to be $\mathcal{O}(10^{2})$ in the volume probed by facilities such as the forthcoming Vera Rubin Observatory Legacy Survey of Space and Time (LSST, \citealt{Ivezic2008}). This indicates the possibility of identifying SMBHBs electromagnetically in meaningful numbers in current and future astronomical datasets.

Indeed, the advent of large, sensitive, and modern time-domain surveys in the past decades has opened us to the possibility of searching for those theoretically predicted signatures amongst a large sample of AGN, making it a promising avenue to identify these intrinsically rare sources. In the past few years alone, more than 100 SMBHB candidates have been proposed in the literature as the results of systematic searches (e.g. \citealt{Graham2015, Graham2015Nat, Charisi2016, Liu2015, Liu2016, Liu2019, Chen2020}) and serendipitous discoveries alike (e.g., \citealt{Shu2020,Hu2020}). While observational searches for SMBHBs can be susceptible to false-positive contamination\footnote{For instance, false positives can result from time-domain searches when the intrinsic temporal variability of ``regular'' AGNs is mistaken for periodicity, especially given limited data length and quality (see \citealt{Vaughan2016}). The high rate of false positives is also suggested by comparing the ensemble GW signal (the gravitational wave background, GWB) implied by the observational sample with the independent PTA constraints on the GWB amplitude (see \citealt{Sesana2018}). }, this can be partially overcome by, e.g., more sensitive observations and longer temporal baselines provided by e.g., the LSST, and it can be expected that EM observatories will continue to expand their roles in the search for SMBHBs.

In addition to the obvious goal of uncovering the elusive SMBHBs, a wealth of astrophysical information about the binary (or binary candidate) can be extracted from EM observations, either through standard observational techniques developed for ``regular'' AGNs, e.g., to estimate the total black hole mass; or applying a specific binary model, e.g., to obtain the mass ratio or orbital period. A remarkable example is the well-known (and currently the best) SMBHB candidate OJ 287 (e.g., \citealt{Sillanpaa1988,Lehto1996,Valtonen2008Nature, Valtonen2016}), where its long-term light curve allows one to obtain an orbital solution and derive parameters including orbital period, primary and secondary BH masses, eccentricity, and even precession rate and primary BH spin. More generally, for binary candidates discovered in time-domain surveys that display apparent periodic variability, the putative period can be determined with time-series analysis techniques, and constraints on other parameters such as orbital inclination and mass ratio may be made by modeling the source light curve under the assumption of the binary model that best describes the possible periodicity (an example is the interpretation of the apparent optical periodicity of PG 1302-102 as relativistic beaming by \citealt{D'Orazio2015}).

This has particularly interesting implications for MMA with SMBHBs. In the context of searching for and studying SMBHBs with the PTA, a ``continuous wave'' (CW) signal from a circular, non-spinning SMBHB can be described by eight independent parameters: sky position (two parameters), chirp mass, GW frequency, inclination, characteristic strain (or luminosity distance), GW polarization angle, and GW phase. With the exception of the last two GW parameters, they could all be determined or constrained by EM (including time-domain) observations. EM information obtained from the source may then be utilized in the search for a CW signal in PTA data and potentially enhance the detectability of the signal, which may otherwise be marginal. Conversely, the PTA GW data estimate the binary parameters in a way that is completely independent from EM observations, and the detection of a second messenger, GWs, from an electromagnetically selected SMBHB candidate and the independent verification of its source parameters would be more robust against false-positive detections and could provide the ultimate confirmation.

The effort to search for GWs in PTA data using the EM information of an SMBHB candidate dates back to the early years of PTA experiments, when \cite{Jenet2004} searched in 7 yr long observations of one pulsar for the timing residuals (that is, difference between the observed and expected pulse arrival times) induced by the proposed SMBHB in the center of the radio galaxy 3C 66B \citep{Sudou2003}, and placed upper limits on its mass and eccentricity from the non-detection of GWs. More recently, \cite{NG11yr3C66B} searched for GWs from 3C 66B in the timing data of 34 pulsars spanning up to 11 yr and placed a constraint on its mass at a level that is almost competitive with EM observations \citep{Iguchi2010}. This is in part attributed to the boosted sensitivity by using EM information as priors: they demonstrated that they were able to gain a factor of two in upper-limit sensitivity by targeting the CW search at the sky position of the source, rather than performing an all-sky search; and an order of magnitude improvement in sensitivity by searching for a signal at the GW frequency suggested by 3C 66B's apparent orbital motion, instead of performing an all-frequency search. This approach is analogous to the ``directed search'' (when sky location is known) and ``targeted search'' (when frequency and other parameters are also known) for the continuous GW signals from rotating, non-axisymmetric neutron stars with LIGO (e.g., \citealt{Aasi2015, Abbott2019}). These searches similarly reduce the computational cost and enhances the sensitivity compared to an all-sky search for an unknown source.

As the PTA sensitivity to nanohertz CWs continues to improve with longer observations and more pulsars, it is poised to put tighter constraints on EM-selected SMBHB candidates, provide more powerful tests of their binary hypothesis, and even make a detection in the near future. Indeed, binary population simulations suggest that a CW detection is possible in the next decade with an improved or next-generation PTA \citep{Rosado2015,Kelley2018SMBHB}, and predications based on observational samples of possible binary hosts are equally promising \citep{Mingarelli2017,Xin2021}. For instance, the latter study suggests that, OJ 287, 3C 66B, and a handful of SMBHB candidates from \cite{Graham2015} (if they are genuine SMBHBs) would be within the detection sensitivity of the next-generation PTA experiment with the Square Kilometre Array \citep{Janssen2015SKA} in the 2030s.

With the dawn of nanohertz GW and multi-messenger astronomy upon us, so is the urgency to systematically and thoroughly understand the PTA's capability as a GW observatory, including its ability to estimate binary parameters and its potential to search for and study the source in a multi-messenger approach. This paper is a step toward that goal, by examining the effects of a multi-messenger search on CW detection and source- parameter estimation using a realistic, mock, data analysis approach. To that end, we inject CW signals of modest GW amplitudes in simulated 11 yr long PTA datasets as proxies of the first detectable CW signals in some future, improved PTA, dataset (Section \ref{sec:inject}). We attempt to recover the signal parameters by performing mock uninformed and EM-informed searches (Sections \ref{sec:blind} and \ref{sec:targeted}). We quantify and compare the results of those searches (Sections \ref{sec:detect} ands \ref{sec:pe}) and discuss implications for future CW searches and SMBHB studies (Section \ref{sec:discuss}). We conclude in Section \ref{sec:conclude}. Throughout the paper, we assume a nine-year WMAP cosmology \citep{WMAP9} and adopt geometrized units with G = c = 1.


\section{Methods} \label{sec:methods}

\subsection{A Simulated PTA and CW Injections}\label{sec:inject}

We first construct a simulated PTA using properties of the pulsars from the NANOGrav 11 yr dataset\footnote{https://data.nanograv.org} \citep{NG11yrdata}. Of the $45$ pulsars that are timed over an 11 yr period, $34$ have time spans of more than three years and were used for NANOGrav's suite of 11-year GW analyses \citep{NG11yrGWB, NG11yr3C66B, NG11yrGa, NG11yrCW, NG11yrGWM}. We additionally remove PSR B1937+21 from the simulated PTA, whose timing residual is known to be characterized by a large amount of red noise above the white noise level (e.g., \citealt{Lam2017,NG11yrdata}). Our simulated PTA thus consists of $33$ pulsars that retain the observing cadence and sky locations (Figure \ref{fig:map}, left panel) of the 11 yr pulsars.

To simulate their timing residuals, we used the \texttt{libstempo} software package\footnote{https://github.com/vallis/libstempo/} and modeled them as a combination of white noise, red noise, and residuals induced by a CW signal. The white noise component represents the nominal TOA measurement uncertainties (\texttt{EFAC=1}), and the red noise component encompasses various effects such as spin noise intrinsic to the pulsar and interstellar medium propagation effects and is modeled as a power law. The power-law model is parameterized by an amplitude $A_{\rm rn}$ and a spectral index $\gamma_{\rm rn}$, which have been measured separately for each pulsar.

To inject a CW signal using the functionality in \texttt{libstempo}, the following signal model parameters are required: \{$\phi, \theta, d_{\rm L}, \mathcal{M}, f_{\rm gw}, i, \Phi_{0}, \psi$\}.

\begin{itemize}

\item The parameters $\phi$ and $\theta$ are the azimuthal angle and polar angle, respectively, and are conventionally used in PTA GW analyses to represent the CW source's position on the sky. In observational astronomy, they correspond to right ascension $\alpha$ and declination $\delta$, respectively: $\phi = \alpha$, $\theta = \pi/2-\delta$. We inject CW sources at two representative sky locations (Figure \ref{fig:map}, left panel): a sensitive sky location where a large number of pulsars are being monitored nearby (R.A. 18$^{\rm h}$, decl. $-15^{\circ}$), and a less sensitive sky location where very few of them are nearby (R.A. 10$^{\rm h}$, decl. $-15^{\circ}$).


\begin{figure*}[ht]
\centering
\epsfig{file=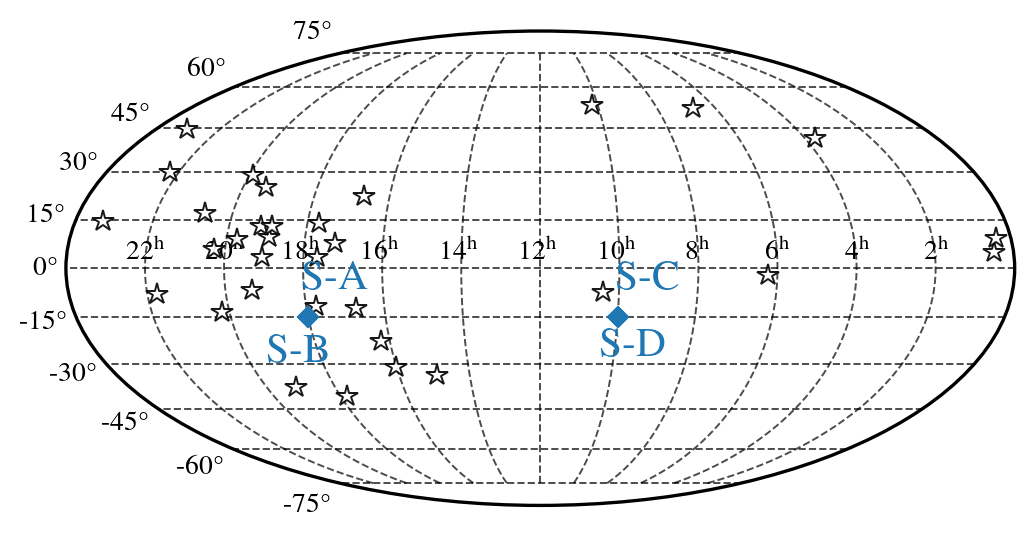,width=0.45\textwidth,clip=}
\epsfig{file=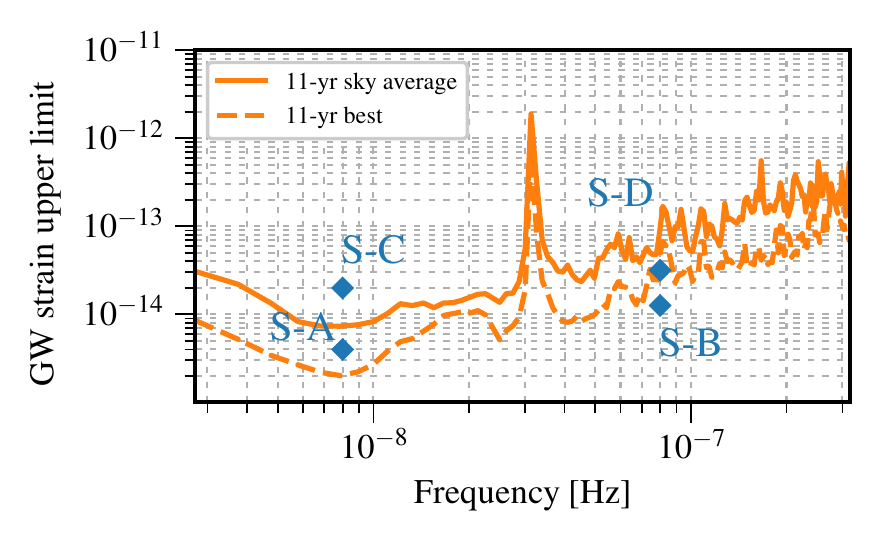,width=0.45\textwidth,clip=}
\caption{Left panel: sky locations of injected CW sources, S-A -- S-D (diamonds). Locations of the pulsars used in the simulated PTA dataset are marked with stars. Right panel: for illustration purposes only, we also show the strain amplitudes of the injected signals compared to the actual 11 yr upper-limit sensitivity on CWs averaged over the sky and at the best sky location (solid and dashed lines, respectively; reproduced from \citealt{NG11yrCW}). Note that S-A and S-B are located near the best sky location.}
\label{fig:map}
\end{figure*}

\begin{table}[ht]
\caption{Injected CW Parameters}
\begin{center}
\begin{tabular}{cccccc}
\hline \hline
Parameter & \multicolumn{5}{c}{Simulation} \\
\hline
 & 0 & A & B & C & D \\ 
\hline
R.A. (hr) & (18)  & 18 & 18 & 10 & 10 \\
Decl. (deg) & (-15) & -15 & -15 & -15 & -15  \\
$f_{\rm gw}$ (nHz) & (8) & 8 & 80 & 8 & 80  \\
$\log \mathcal{M}$ ($M_{\odot}$) & (9.4) & 9.4 &  9.4 & 9.5 & 9.5 \\
$z$  & (0.06) & 0.06  & 0.08 & 0.02 & 0.05 \\
\hline
$\log h_{0}$ & \nodata & -14.4 & -13.9 & -13.7 & -13.5  \\
\hline \hline
\end{tabular}
\end{center}
\tablecomments{For all simulations: $q$=0.8, $i$=$\pi/2$, $\Phi_{0}$ = 2.0, $\psi$ = 2.0. Additionally, S-0 does not contain an injected CW signal, but the search was performed using the parameters of S-A (shown in parentheses).}
\label{tab:inject}
\end{table}

\begin{table}[ht]
\caption{$\mathcal{F}_{\rm p}$ Values and SNRs of different Realizations of S-A -- S-D}
\begin{center}
\begin{tabular}{ccc}
\hline \hline
 & $\mathcal{F}_{\rm p}$ & SNR \\
 \hline
S-A1 & 55 & 6.6 \\ 
S-A2 & 45 & 4.9  \\ 
\hline
S-B1 & 40 & 3.7 \\ 
S-B2 & 44 & 4.7 \\ 
\hline
S-C1 & 42 & 4.2  \\ 
S-C2 & 43 & 4.5 \\ 
\hline
S-D1 & 30 & $<1$ \\ 
S-D2 & 32 &  $<1$ \\ 
\hline \hline
\end{tabular}
\end{center}
\label{tab:realizations}
\end{table}


\item The parameter $d_{\rm L}$ is the luminosity distance of the CW source. Assuming a cosmology, this can be obtained observationally by measuring the redshift $z$ of the host galaxy of the SMBHB, or the AGN for an active SMBHB. We injected mock CW sources in the very nearby universe at $z<0.1$ to mimic the detectable signals that are already present in the respective simulated datasets.

\item The parameter $f_{\rm gw}$ is the GW frequency and is related to the orbital frequency of the SMBHB: $f_{\rm gw} = 2f_{\rm orb}$ (assuming a circular orbit). We inject CW signals at two different frequencies (Figure \ref{fig:map}; right panel): $f_{\rm gw}=8$ nHz, which is approximately the most sensitive frequency in the 11yr dataset; and $f_{\rm gw}=80$ nHz, which is motivated by the observational search for AGN periodicity: in time-series data with a baseline of a few years (which is typical for current time-domain surveys), at least a few cycles are required before claiming a periodic detection, hence making the EM search more sensitive to periods of approximately months. Furthermore, the stochastic GW background (GWB), on which the CW signal would be superimposed, has a canonical power-law spectral shape and thus has a lower amplitude at high GW frequencies, potentially making the search for a CW source emitting at a high GW frequency more fruitful.

\item $\mathcal{M}$ is the (redshifted) ``chirp mass'' of the SMBHB and is a combination of the component masses: $\mathcal{M} \equiv (m_{1}m_{2})^{3/5}/(m_{1}+m_{2})^{1/5}$. Since our injected CW sources are in the very local universe, we make the approximation that the rest-frame chirp mass $\mathcal{M}_{r} \approx \mathcal{M}$. The amplitude of the GW signal is then related to $d_{\rm L}$, $f_{\rm gw}$, and $\mathcal{M}$: 

\begin{equation}
h_{0} = \frac{2\mathcal{M}^{5/3}(\pi f_{\rm gw})^{2/3}}{d_{\rm L}},
\label{eqn:h}
\end{equation}

In our analyses, we parameterize the signal strength in terms of the total mass $M_{\rm tot} = m_{1}+m_{2}$ and the mass ratio $q \equiv m_{2}/m_{1}\le1$ rather than the chirp mass because those parameters are more closely tied to EM observations. Observationally, the total mass of the system can be measured via standard methods such as single-epoch virial black hole mass estimation (assuming the method is still valid for close binary systems), which suffers from an $\sim$$0.3$ dex systematic uncertainty (e.g., \citealt{Shen2013} and references therein). In this study, we assign astrophysically reasonable black hole masses to the injected CW sources: $\log\mathcal{M/M_{\odot}}=9.4-9.5$. The mass ratio of a binary system cannot always be directly determined from the light curve of a periodically varying SMBHB (see Section \ref{sec:discuss} for further discussion), and we keep the mass ratio fixed at $q=0.8$ in all injections for simplicity.

\item The parameter $i$ is the inclination of the binary orbit and could be observationally constrained, including through light curve fitting. We keep the inclination fixed at $i = \pi/2$ (which corresponds to an edge-on binary) for all injections for simplicity. We note, that while an edge-on or near edge-on orientation is optimal for EM observers in the cases of relativistic beaming and binary self-lensing, it is the least optimal for GW detection because it induces the smallest pulsar timing residuals.

\item The parameters $\Phi_{0}$ and $\psi$ are the initial orbital phase and the GW polarization angle in units of radians, respectively. As they cannot be measured from EM observations, we keep them fixed at $\Phi_{0}=2.0$ and $\psi=2.0$ in all injections for simplicity.

\end{itemize} 

The parameter values for each injected CW signal are summarized in Table \ref{tab:inject}. Additionally, we have included the so-called ``pulsar term'' in the injected signal, which is the signal induced at the pulsar (whereas ``Earth term'' is the signal induced at the Earth). In contrast to the alternative approach that uses a signal model that only contain the Earth term, our analysis is sensitive to the evolution, or ``chirping,'' of the signal, and hence the chirp mass, as GWs pass through the pulsar and the Earth.

For each simulation (Simulations A-D), we first calculate the GW strain $h_{0}$ from the binary parameters (Figure \ref{fig:map}, right panel; Table \ref{tab:inject}). We have generated two realizations for each simulation, denoted S-X1 and S-X2, in order to study the potential effect of pulsar white and red noise realizations on the CW search. For each realization, we compute the $\mathcal{F}$-statistic, which is the log-likelihood ratio maximized over the so-called extrinsic parameters of the source and is used as a detection statistic for GW searches \citep{Jaranowski1998,Cornish2007,Babak2012}. In the context of this work, we evaluate the pulsar-term $\mathcal{F}$-statistic ($\mathcal{F}_{\rm p}$, \citealt{Ellis2012}) at the injected GW frequency. We additionally compute the corresponding SNR, since $2\mathcal{F}_{\rm p}$ follows a $\chi^{2}$ distribution with $2N$ degrees of freedom, where $N$ is the number of the pulsars: SNR = $\sqrt{2\mathcal{F}_{\rm p}-2N}$. We show the $\mathcal{F}_{\rm p}$ and SNR of each realization of S-A--S-D in Table \ref{tab:realizations}. The different mock datasets containing the same injected CW signal differ slightly in $\mathcal{F}_{\rm p}$ and SNR, as expected given their different noise realizations.

As we see in Table \ref{tab:inject} and Table \ref{tab:realizations}, Simulation A has the highest $\mathcal{F}_{\rm p}$ statistic or SNR despite having the lowest nominal GW amplitude, which can be expected as it is emitting at a GW frequency favorable to the PTA. Conversely, Simulation D has the largest GW amplitude but has the lowest detection statistic or SNR, due to its GW frequency to which the NANOGrav PTA is less sensitive. Overall, Simulations A-D represent CW signals of a range of parameters and are capable of serving as test cases in this proof-of-concept study to investigate any effects of sky location and GW frequency on CW searches. Their comparable SNRs further permit reasonable comparisons between the mock CW search results. Moreover, their modest SNRs are suitable proxies for the first CW signals in some future PTA dataset, which are expected to be weak.

We note that, in real PTA data, any CWs detectable as discrete signals would be superimposed on the stochastic GWB, which is the superposition of GW signals from a cosmological population of SMBHBs, and it is probable that multiple CW signals may be detectable simultaneously above the GWB. However, since the 11-year NANOGrav analysis does not show evidence for the GWB \citep{NG11yrGWB}, which is expected to emerge in a dataset spanning at least 15 years \citep{Taylor2016,Pol2021}, we do not inject a GWB in our simulated PTA datasets. Additionally, the joint search for one or more CW signals in addition to a GWB \citep{Becsy2020} requires  techniques such as a trans-dimensional Reversible Jump MCMC sampler (e.g., \citealt{Green1995}) to explore a parameter space whose dimensionality is not fixed, which is beyond the scope of this work. For these reasons, we only inject one CW source in a given simulated PTA dataset.

We further note that we have based our simulated PTA on the NANOGrav 11 yr dataset, instead of the more recent 12.5 yr data set \citep{NG12p5yrdata}. Since a red noise process common among the 12.5 yr pulsars is present in this dataset but whose origin cannot yet be determined \citep{NG12p5GWB}, we do not utilize this dataset in our analysis. Also, an 11 yr like simulated dataset permits reasonable comparisons with NANOGrav results from the latest CW analysis, which was performed with the 11 yr dataset \citep{NG11yrCW}. Furthermore, we argue that the main results from this proof-of-concept work should not strongly depend on the length and size of the particular dataset, since the SNRs of the injected CW signals are relative to the sensitivity of the dataset. That said, extending this work to a later dataset, either actual or projected, would be of interest to future work.

Finally, we generate two additional mock datasets with no injected CW signals (S-0). In Sections \ref{sec:blind} and \ref{sec:targeted}, we will treat S-0 in the same manner as S-A when performing uninformed and targeted searches (Table \ref{tab:inject}) and will compare and contrast the search results. This is motivated by the high false-positive rate of SMBHB searches (see, e.g., the footnote in Section \ref{sec:intro} for a brief discussion), and searching for a CW signal in S-0 thus represents the case where a CW search is performed using the information of an EM candidate which turns out to be a false positive.


\subsection{Mock uninformed searches} \label{sec:blind}

For each injection, we first perform a ``blind'' search, where the MCMC sampler is free to explore the prior ranges of all parameters (Table \ref{tab:blind_prior}). This procedure thus mimics an unguided and uninformed search for a CW signal in PTA data. We perform these mock searches using the NANOGrav GW detection software package \texttt{enterprise}\footnote{https://github.com/nanograv/enterprise}. As we largely adopt the same data analysis methods as the previous CW analyses (modified for a simulated PTA dataset), we refer the reader to \cite{Ellis2012} or \cite{NG11yrCW} for a detailed description of the CW signal model and search methods. To explore the parameter space, we use \texttt{PTMCMCSampler}\footnote{https://github.com/jellis18/PTMCMCSampler} \citep{PTMCMC}, with a geometrically spaced temperature ladder for parallel tempering (e.g., \citealt{Earl2005}). Each chain is then sampled for $(1-2)$$\times10^{6}$ iterations and individually inspected for convergence using the Geweke diagnostic \citep{Geweke1992}. 



\begin{table}[ht]
\caption{Priors in uninformed searches}
\begin{center}
\begin{tabular}{cccc}
\hline \hline
Parameter & Prior & Range & Symbol in Table \ref{tab:note} \\
\hline
$\phi$ & Uniform & [0, 2$\pi$] & \xmark \\
$\cos\theta$ & Uniform & [-1.0, 1.0] & \xmark \\
$\log d_{\rm L}$ & Uniform & [-2.0, 4.0] & \xmark \\
$\log M_{\rm tot}$ & Uniform & [6.0, 12.0] & \xmark \\
$\log f_{\rm gw}$ & Uniform & [-9.0, -7.0] & \xmark \\
$q$ & Uniform & [0.0001,1.0] & \nodata \\
$\Phi_{0}$ & Uniform & [0, $\pi$] & \nodata \\
$\psi$ & Uniform & [0, $\pi$] & \nodata \\
$\cos i$ & Uniform & [-1.0, 1.0] & \nodata \\
\hline \hline
\end{tabular}
\end{center}
\tablecomments{$q$, $\cos i$, $\Phi_{0}$, and $\psi$ are always searched over their respective priors and are hence omitted in Table \ref{tab:note}.}
\label{tab:blind_prior}
\end{table}

\begin{table}[ht]
\caption{Priors in targeted searches}
\begin{center}
\begin{tabular}{ccc}
\hline \hline
Parameter & Prior & Symbol in Table \ref{tab:note} \\
\hline
$\phi$ & Fixed & \cmark \\
$\cos \theta$ &  Fixed  & \cmark \\
$\log d_{\rm L}$ & Fixed  & \cmark  \\
$\log M_{\rm tot}$ & Normal & 0.3 dex \\
$\log f_{\rm gw}$ & Uniform & $\times$4  \\
\hline \hline
\end{tabular}
\end{center}
\label{tab:tgt_prior}
\end{table}

\begin{table*}[ht]
\caption{Names of uninformed and targeted searches in Figures \ref{fig:tgt_sim31}-\ref{fig:tgt_sim98} }
\begin{center}
\begin{tabular}{ccccc}
\hline \hline
Parameter & \multicolumn{4}{c}{Name} \\
\hline
 & {\it uninformed} & {\it ra\_dec\_z} & {\it ra\_dec\_z\_dmtot} & {\it ra\_dec\_z\_dmtot\_dfgw} \\
\hline
$\phi$ & \xmark & \cmark & \cmark & \cmark \\
$\cos \theta$ & \xmark & \cmark & \cmark & \cmark \\
$\log d_{\rm L}$ & \xmark & \cmark & \cmark & \cmark \\
$\log M_{\rm tot}$ & \xmark & \xmark & 0.3 dex & 0.3 dex \\
$\log f_{\rm gw}$ & \xmark & \xmark & \xmark & $\times$4 \\
\hline \hline
\end{tabular}
\end{center}
\label{tab:note}
\end{table*}


\subsection{Mock targeted searches} \label{sec:targeted}

To study the effects of applying EM-informed priors on the detection and parameter estimation of a CW signal, we perform three separate searches for the same injected signal in a progressive scheme:

\begin{itemize}

\item {\it ra\_dec\_z}: We search for the signal at its injected sky location and luminosity distance. This mimics the search for a CW signal in the PTA dataset which is prompted by the detection (or observation) of its EM counterpart (or candidate). In this scenario, we assume the SMBHB's location and redshift can be measured exactly, which is a reasonable assumption for EM observations of a source.

\item {\it ra\_dec\_z\_dmtot}: We search for the signal at its injected sky location, luminosity distance, and in addition, we take into account a 0.3 dex systematic uncertainty on the black hole mass measurement. In practice, this is implemented through adopting a normal prior for $\log M_{\rm tot}$ which is centered at the injected value. We note that, in practice, this scenario almost always contains the previous case; in other words, if an EM counterpart can be identified at all, its black hole mass can usually be estimated, and therefore in practice, {\it ra\_dec\_z} is never a stand-alone case for a targeted search. However, here we study both cases in order to isolate the effects of knowing the black hole mass on the targeted search.

\item {\it ra\_dec\_z\_dmtot\_dfgw}: This represents the case where the SMBHB is detected electromagnetically as a periodically varying AGN,  where the observed period is interpreted as the imprint of the binary orbital period. In addition to the priors above, we take into account the uncertainty of the GW frequency, which is motivated by factors including measurement uncertainties and being agnostic about how the observed variability period translates to the binary orbital period (which may have a binary model and/or mass-ratio dependence, see Section \ref{sec:discuss} for details). In practice, we apply a uniform prior to $\log f_{\rm gw}$, while allowing a factor of two uncertainty around the injected value, i.e., a total of a factor of four uncertainty on $f_{\rm gw}$.

\end{itemize}

The priors applied in targeted searches are listed in Table \ref{tab:tgt_prior}, and we summarize the uninformed search and all targeted searches in Table \ref{tab:note}.


\section{Results and Discussion} \label{sec:results}

\subsection{Detectability}\label{sec:detect}

We first investigate the effects of a targeted search on the detectability of the CW signal. To this aim, we compute the Bayes factor for the presence of a CW signal. We use the Savage-Dickey formula \citep{Dickey1971}, which can be used to compute the Bayes factor $\mathcal{B}_{10}$ for two nested models $\mathcal{H}_{1}$ and $\mathcal{H}_{0}$. In this case, $\mathcal{H}_{0}$ is a model with pulsar noise only, and $\mathcal{H}_{1}$ is a model with noise plus a CW signal, which is equivalent to noise only for $h_{0}=0$. For more details, see Appendix \ref{append:a}

In Figure \ref{fig:sim_bf}, we show the evolution of the Bayes factor from an uninformed search to targeted searches. In {\it uninformed}, all Bayes factors $\mathcal{B}_{10}<1$, indicating that no evidence for the CW signal can be found. This is expected from the low SNRs of the signals. In {\it ra\_dec\_z}, $\mathcal{B}_{10}$ has increased slightly, but stays below the nominal threshold of $\mathcal{B}_{10} =1$, or barely above it. Thus, knowing the source's sky location does not appear to significantly improve the detectability of the source. However, by additionally knowing the the black hole mass ({\it ra\_dec\_dmtot}), the increases in $\mathcal{B}_{10}$ are more significant (except for S-C1); in particular, $\mathcal{B}_{10}$ in S-A and S-B have increased by at least a factor of a few. In {\it ra\_dec\_dmtot\_dfgw}, where the frequency is also known, $\mathcal{B}_{10}$ increases even further in some cases. S-A and S-B have the larger final $\mathcal{B}_{10}$ values ($\sim 5-10$) among the four injections. While a Bayes factor of $\sim 10$ is shy of being able to claim a confident detection, it is noteworthy that evidence for the (weak) CW signal can strengthen by at least a factor of a few via the EM observation of the source.

Additionally, the overall trend described above appears to hold for both realizations, albeit with mild variation between realizations. The variation is likely the result of the MCMC analysis or the noise realization, rather than a simple correlation with SNR of the source. For example, S-A2 contains the same CW signal as S-A1 and has a slighter lower SNR, but has the highest final $\mathcal{B}_{10}$. Similarly, S-C1 and S-C2 have almost identical SNRs, but an overall increase in $\mathcal{B}_{10}$ is seen in S-C2, while a slight decrease is observed in S-C1.

Despite the individual variations, the evolution trends of S-A and S-B appear to be separated from those of S-C and S-D (last panel in Figure \ref{fig:sim_bf}). This is likely due to the fact that S-A and S-B are located at a sensitive sky location, which is favorable for CW searches originally (see, e.g., \citealt{NG11yrCW}) and benefits from EM information even further. However, we note that the Savage Dickey approximation starts to break down for stronger detections, and therefore we advise the reader to take the $\mathcal{B}_{10}$ values (and their error bars) of S-B2 with a grain of salt.


\pagebreak
\begin{figure*}[ht]
\centering
\epsfig{file=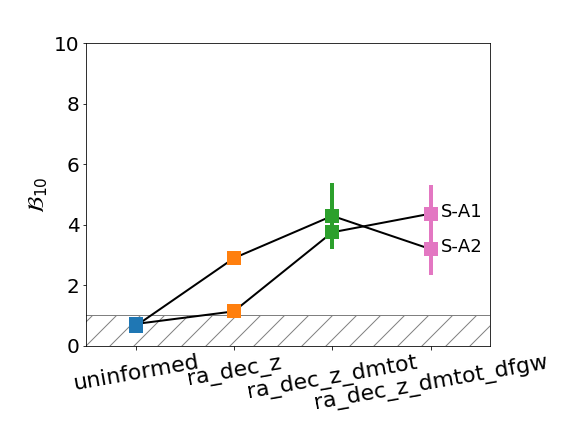,width=0.45\textwidth,clip=}
\epsfig{file=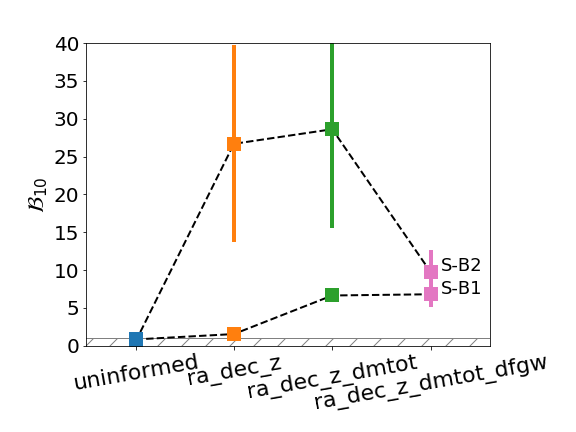,width=0.45\textwidth,clip=}
\epsfig{file=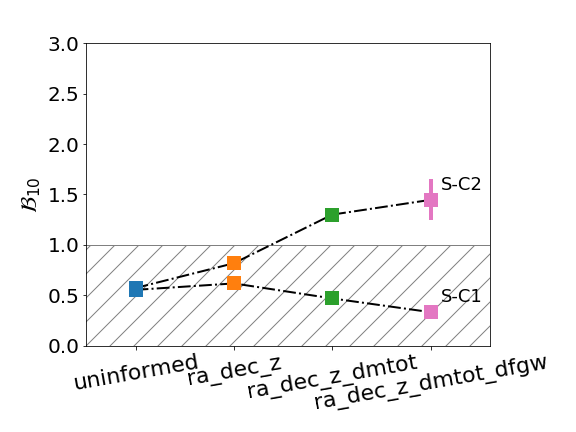,width=0.45\textwidth,clip=}
\epsfig{file=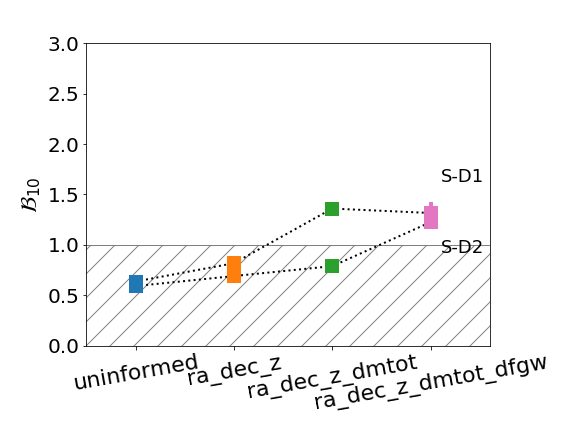,width=0.45\textwidth,clip=}
\epsfig{file=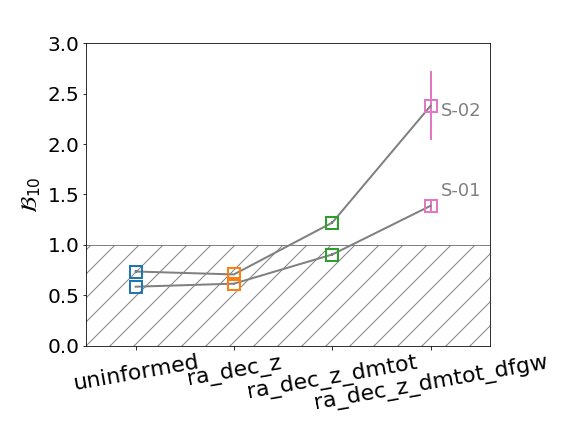,width=0.45\textwidth,clip=}
\epsfig{file=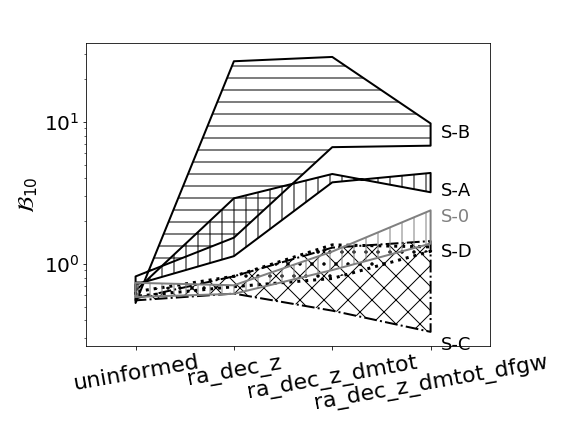,width=0.45\textwidth,clip=}
\caption{
The first five panels show the mean Bayes factors $\mathcal{B}_{10}$ for the presence of a CW signal in {\it uninformed}, {\it ra\_dec\_z}, {\it ra\_dec\_dmtot}, {\it ra\_dec\_dmtot\_dfgw} (blue, orange, green, purple squares, respectively) for the two realizations of S-0 -- S-D. The error bars represent the uncertainties of $\mathcal{B}_{10}$ (see Appendix \ref{append:a}). Note that some of the error bars are not visible due to their small size. The grey hashed region below the nominal detection threshold of $\mathcal{B}_{10}=1$ is shown for visual reference. Note the different ranges of the y-axes.
The last panel compares the evolution of the $\mathcal{B}_{10}$ shown in the first five panels: S-A -- vertical hashed; S-B -- horizontal hashed; S-C -- crossed hashed; S-D -- dotted hashed; S-0 -- grey vertical hashed. The upper and lower bounds are given by the respective mean $\mathcal{B}_{10}$ of the two realizations S-X1 and S-X2; uncertainties in individual realizations are omitted for clarity. The y-axis is shown in log-scale for clarity.
}
\label{fig:sim_bf}
\end{figure*}

Furthermore, comparing the trends of S-A versus S-B, the latter appears to be more detectable than the former: both realizations, S-B1 and S-B2, have higher final $\mathcal{B}_{10}$ than S-A. This is despite the fact that S-B has a quieter signal than S-A. We speculate that this is because S-B is emitting at a higher GW frequency than S-A, so that the CW signal is observed for more cycles over the same data length. While higher frequencies are conventionally considered ``less sensitive'' frequencies in terms of the GW amplitude, our results suggest that sources emitting at high frequencies could be as detectable as (or even more detectable than) those emitting at ``sensitive'' frequencies. This is encouraging for the possibility of strong synergies with EM observations (see Section \ref{sec:discuss} for further discussion). However, we are unable to observe the same dependence on frequency for S-C versus S-D, as their improvements in $\mathcal{B}_{10}$ are much weaker.

We caution that these trends ought to be confirmed with a larger number of realizations and parameters than we have included here, especially given the aforementioned variation between different noise realizations. Unfortunately, it is prohibitively expensive to perform MCMC analysis on a statistically meaningful number of CW searches, for which a more robust and efficient detection pipeline is still needed. Therefore, while our current analyses suggest a dependence of CW detectability on sky location and GW frequency, at present we are not able to fully investigate the effect of different noise realizations on this trend. 


\subsection{False-positive Detections}\label{sec:s0}

We then proceed to investigate whether the above trend applies to S-0, which does not contain an injected CW signal. As we show in the bottom left panel in Figure \ref{fig:sim_bf}, both realizations have slightly increased Bayes factors as a result of targeted searches. Therefore, applying (incorrect) EM priors appears to also risk increasing the false alarm probability. Additionally, the final Bayes factor in {\it ra\_dec\_z\_dmtot\_dfgw} is $\sim2$, which is comparable to that of S-C or S-D. Thus in actual searches, it would be difficult to distinguish a false-positive detection from a low SNR source, if the Bayes factor and its increase are interpreted at face value.

However, it is important to note that the final Bayes factor is significantly below that of S-A, whose source parameters were used in the search for a signal in S-0. Therefore, we argue that the trend of increased $\mathcal{B}_{10}$ of true CW signals (and S-A in particular) cannot simply be the result of possibly increased false alarm probability due to the EM-informed search. Instead, this indicates that a false-positive detection is less likely to be mistaken for a true CW source if it is located at a favorable sky location.

More importantly, this emphasizes that the Bayes factor is not a ``one size fits all'' number to quantify the evidence for a signal, nor is there an absolute threshold between detection and non-detection. In actual CW searches (and especially targeted searches), the resulting Bayes factor should be interpreted in the appropriate context, for example by performing a false alarm analysis similar to this one. Finally, the ``detection'' of a CW signal is only part of the story. EM observations carry with them the parameters of the possible binary, which should be used for cross validation with GW searches in order to further reduce possible false alarms. In the next section, we will investigate the ability of the PTA to estimate, and independently verify, binary parameters in targeted searches.

\subsection{Parameter Estimation}\label{sec:pe}

\begin{figure*}
\centering

\epsfig{file=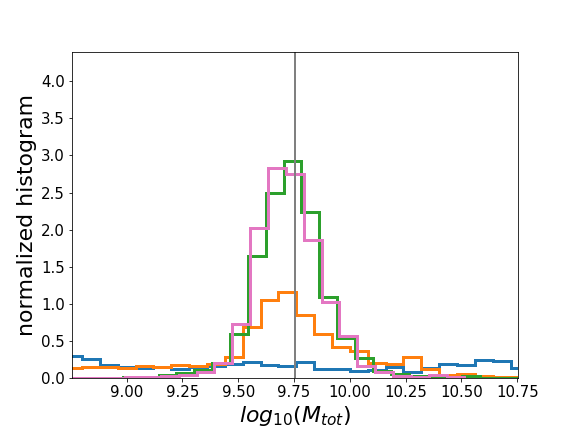,width=0.32\textwidth,clip=}
\epsfig{file=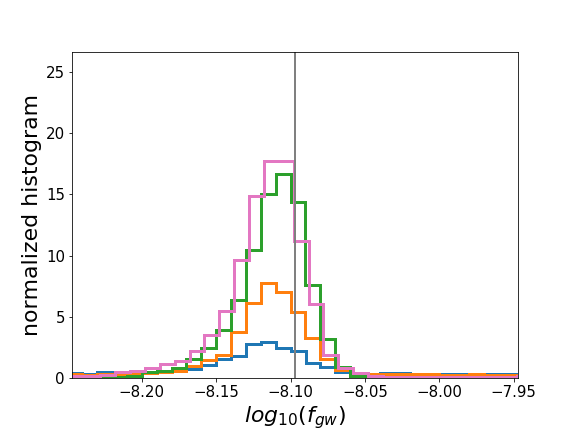,width=0.32\textwidth,clip=}
\epsfig{file=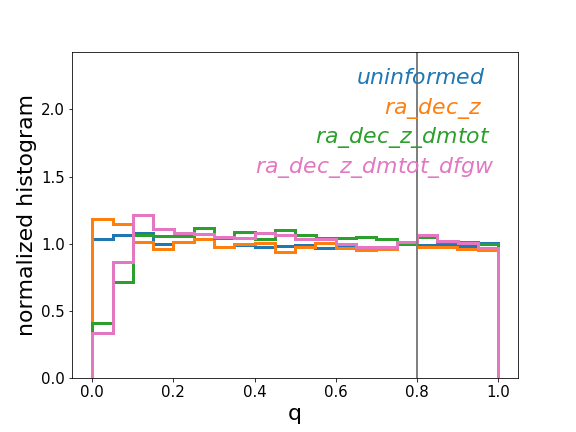,width=0.32\textwidth,clip=}

\epsfig{file=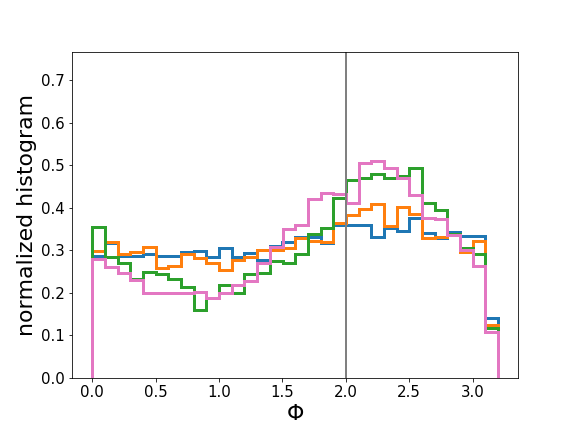,width=0.32\textwidth,clip=}
\epsfig{file=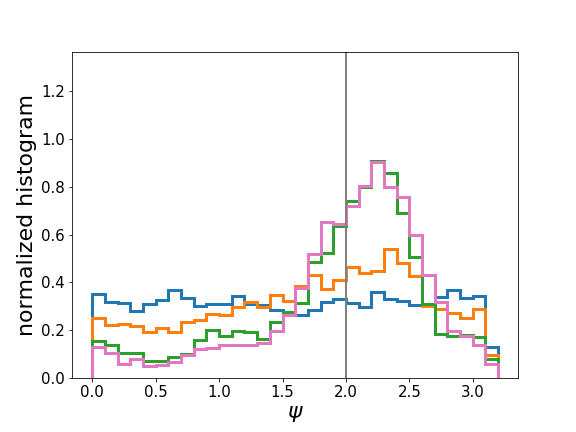,width=0.32\textwidth,clip=}
\epsfig{file=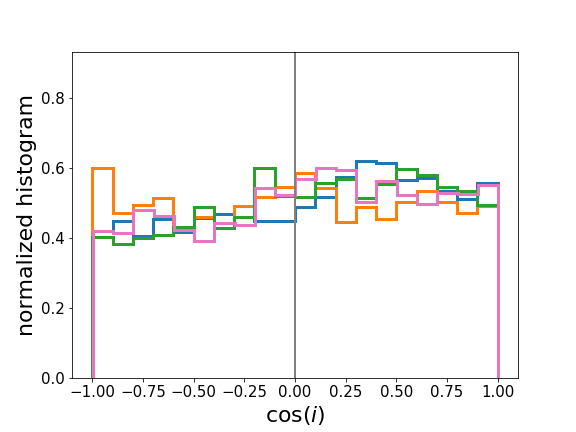,width=0.32\textwidth,clip=}

\caption{An uninformed search and three types of targeted searches are performed for Simulation A1 in order to obtain the posterior distributions for \{$M_{\rm tot}, f_{\rm gw}, q, \Phi, \psi, i$\}. The blue, orange, green, and purple colors correspond to {\it uninformed}, {\it ra\_dec\_z}, {\it ra\_dec\_dmtot}, {\it ra\_dec\_dmtot\_dfgw}, respectively (see Table \ref{tab:note} and text for details). The areas under the histograms have been normalized to one. The injected value is marked with a grey solid line. 
The injected CW signal in S-A1 is located at a favorable sky location and has a $f_{\rm gw}$ of 8 nHz.
}
\label{fig:tgt_sim31}
\end{figure*}

\begin{figure*}
\centering

\epsfig{file=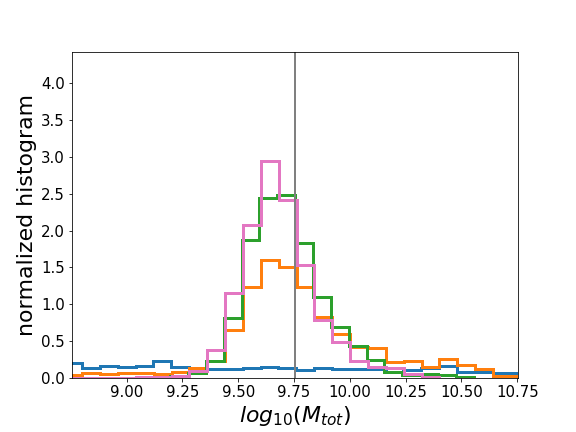,width=0.32\textwidth,clip=}
\epsfig{file=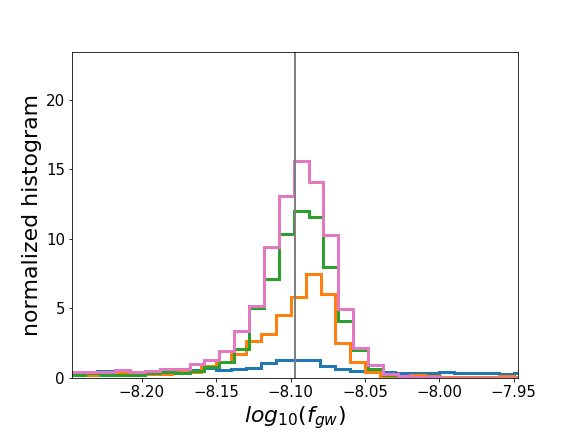,width=0.32\textwidth,clip=}
\epsfig{file=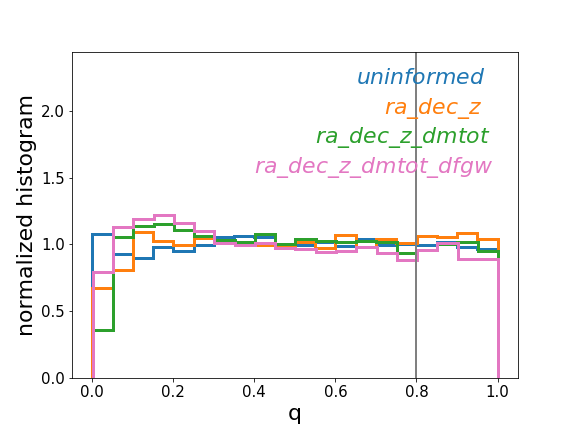,width=0.32\textwidth,clip=}

\epsfig{file=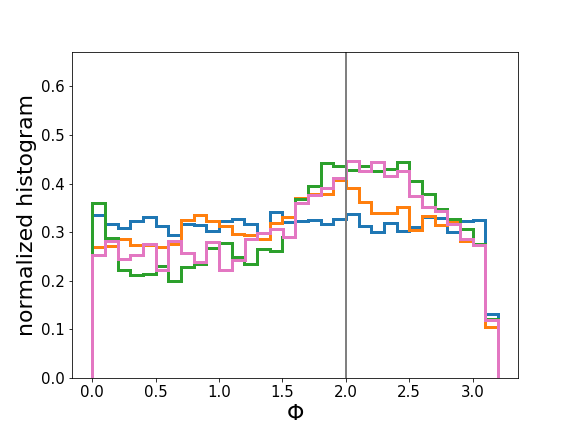,width=0.32\textwidth,clip=}
\epsfig{file=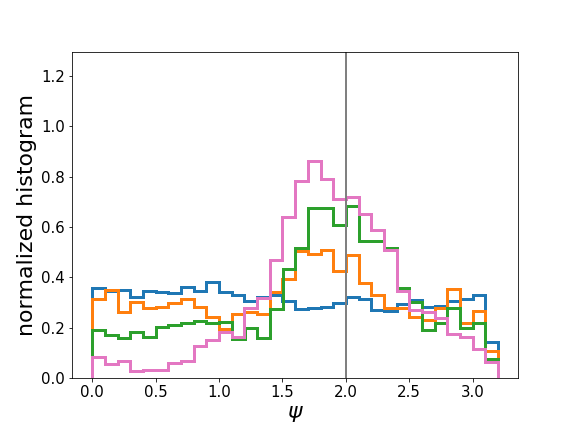,width=0.32\textwidth,clip=}
\epsfig{file=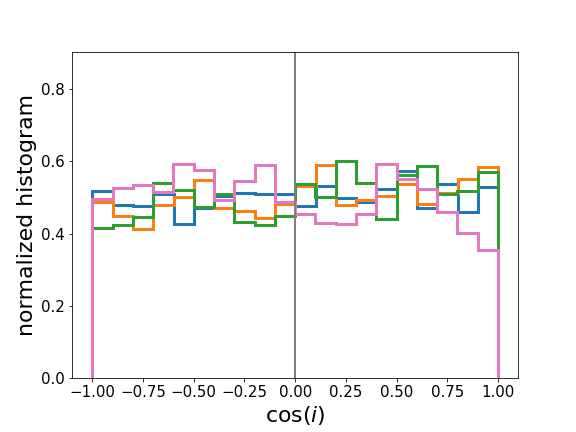,width=0.32\textwidth,clip=}

\caption{Same as Figure \ref{fig:tgt_sim31}, but for S-A2.}
\label{fig:tgt_sim36}
\end{figure*}

\begin{figure*}
\centering

\epsfig{file=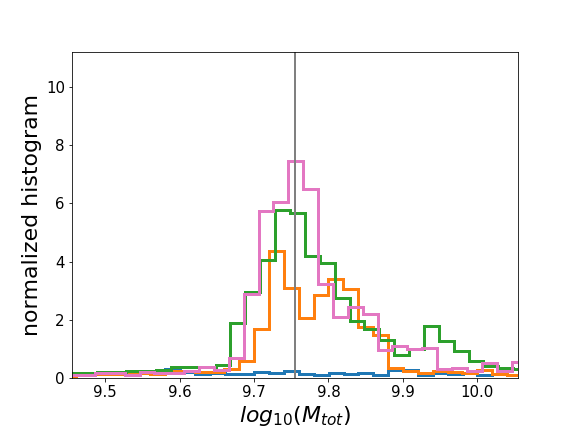,width=0.32\textwidth,clip=}
\epsfig{file=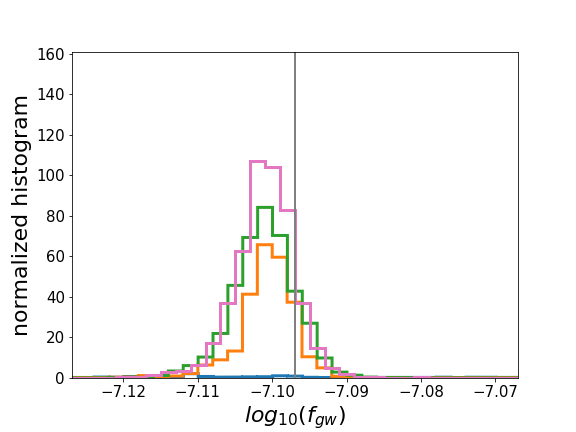,width=0.32\textwidth,clip=}
\epsfig{file=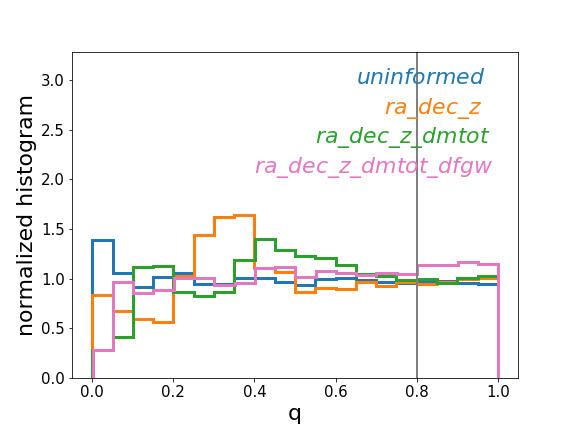,width=0.32\textwidth,clip=}

\epsfig{file=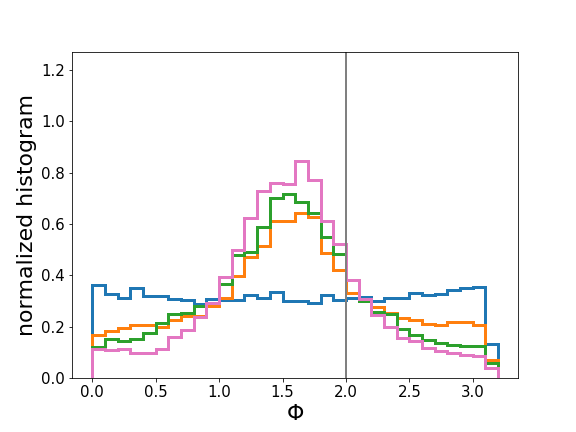,width=0.32\textwidth,clip=}
\epsfig{file=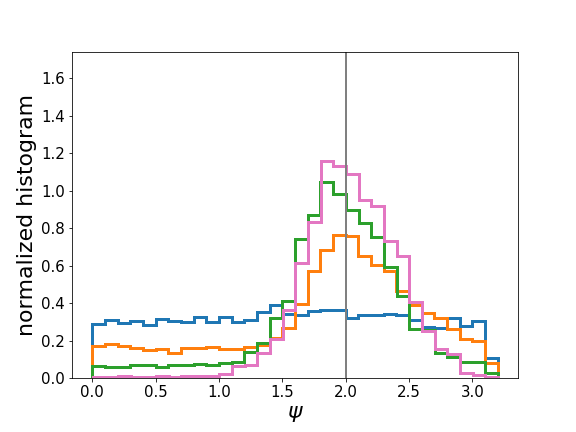,width=0.32\textwidth,clip=}
\epsfig{file=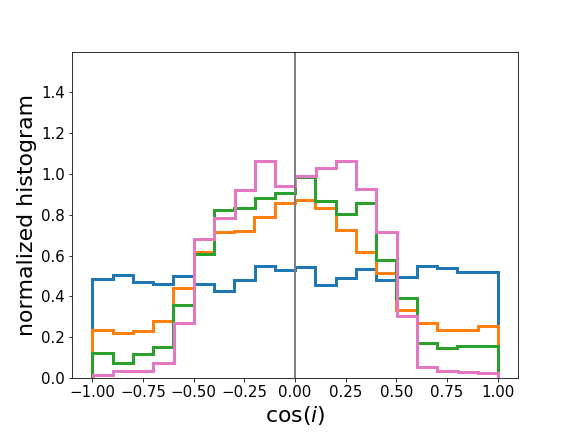,width=0.32\textwidth,clip=}

\caption{Same as Figure \ref{fig:tgt_sim31}, but for Simulation B1. The injected CW signal in S-B1 is located at a favorable sky location and has a $f_{\rm gw}$ of 80 nHz.}
\label{fig:tgt_sim52}
\end{figure*}

\begin{figure*}
\centering

\epsfig{file=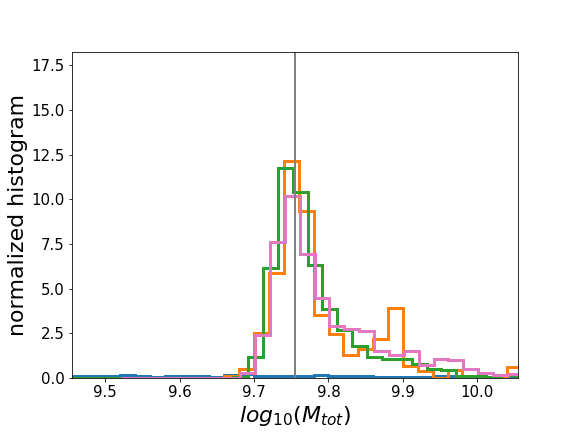,width=0.32\textwidth,clip=}
\epsfig{file=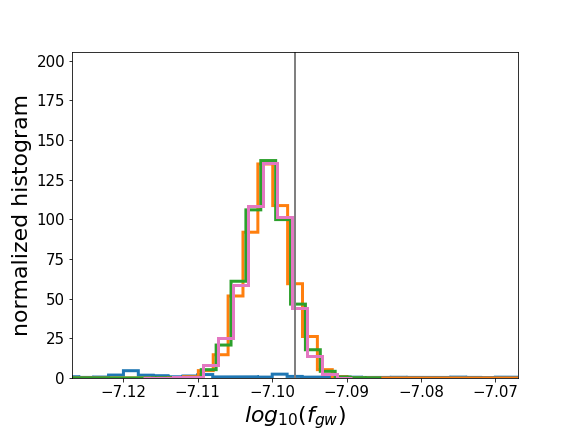,width=0.32\textwidth,clip=}
\epsfig{file=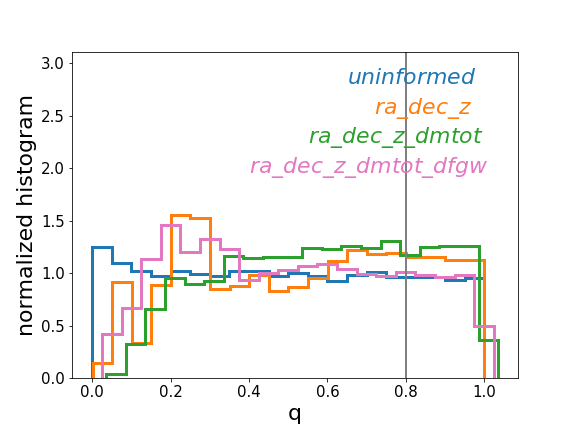,width=0.32\textwidth,clip=}

\epsfig{file=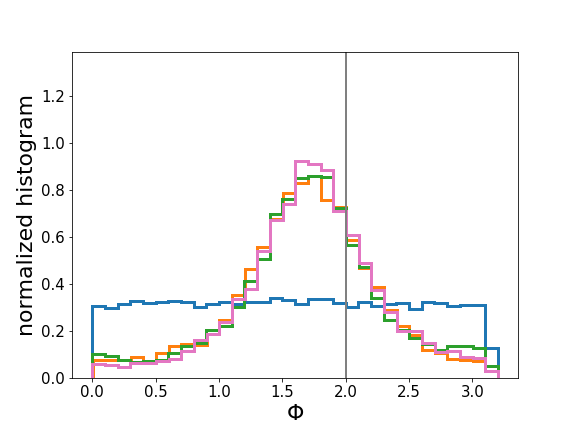,width=0.32\textwidth,clip=}
\epsfig{file=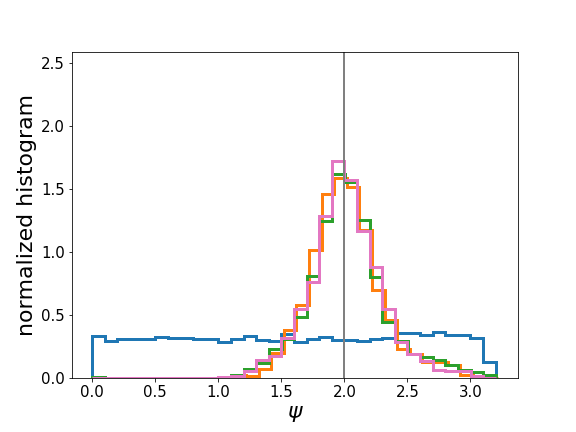,width=0.32\textwidth,clip=}
\epsfig{file=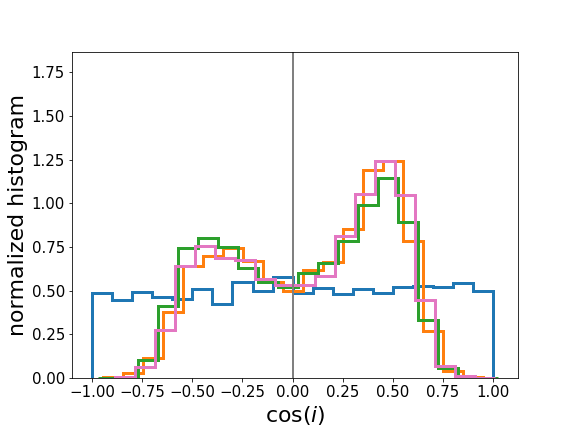,width=0.32\textwidth,clip=}

\caption{Same as Figure \ref{fig:tgt_sim52}, but for S-B2.}
\label{fig:tgt_sim58}
\end{figure*}

\begin{figure*}
\centering

\epsfig{file=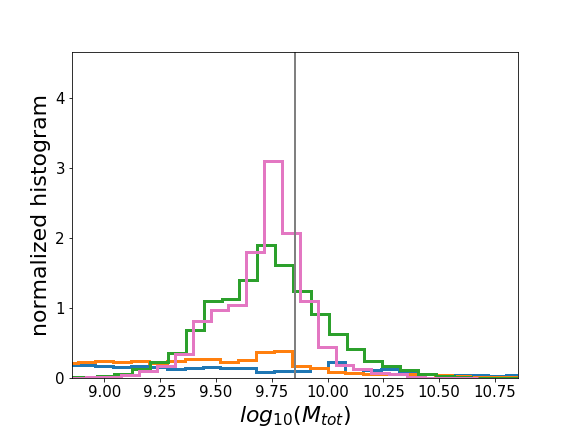,width=0.32\textwidth,clip=}
\epsfig{file=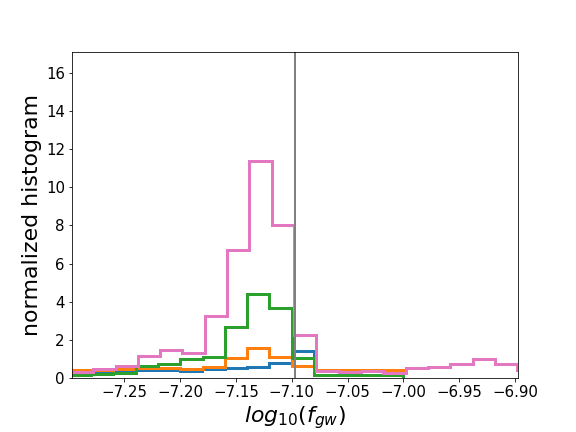,width=0.32\textwidth,clip=}
\epsfig{file=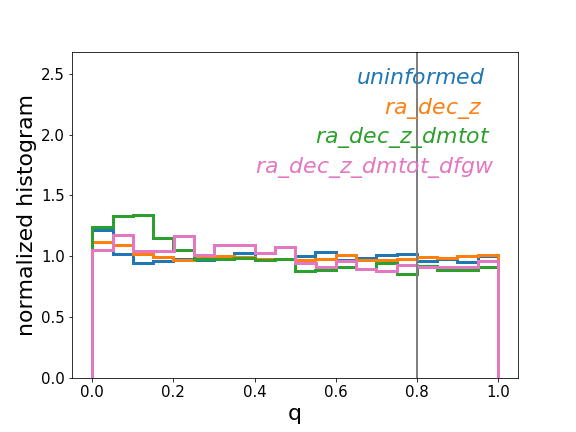,width=0.32\textwidth,clip=}

\epsfig{file=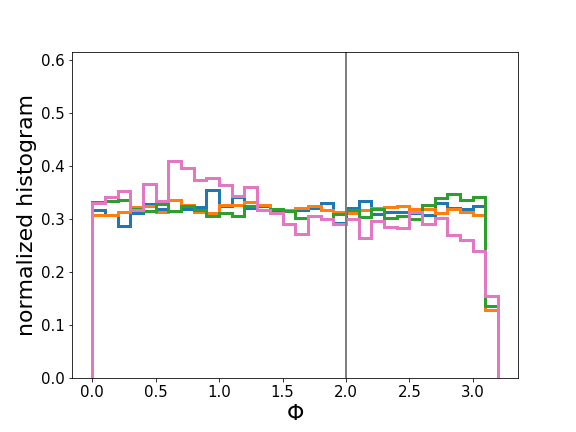,width=0.32\textwidth,clip=}
\epsfig{file=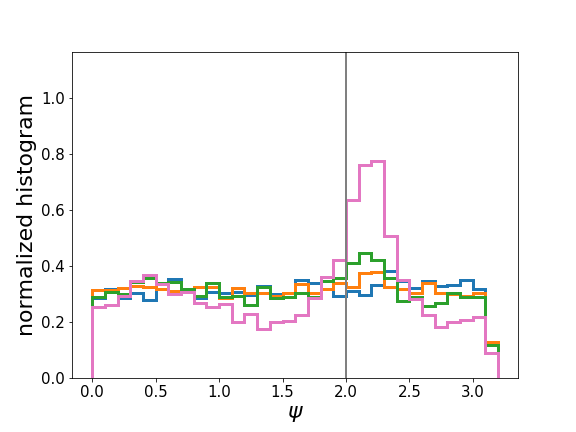,width=0.32\textwidth,clip=}
\epsfig{file=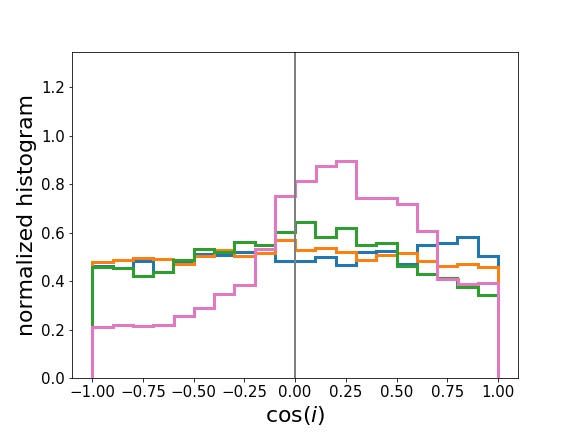,width=0.32\textwidth,clip=}

\caption{Same as Figure \ref{fig:tgt_sim31}, but for Simulation D1. The injected CW signal in S-D1 is located at a less favorable sky location and has a $f_{\rm gw}$ of 80 nHz.}
\label{fig:tgt_sim91}
\end{figure*}

\begin{figure*}
\centering

\epsfig{file=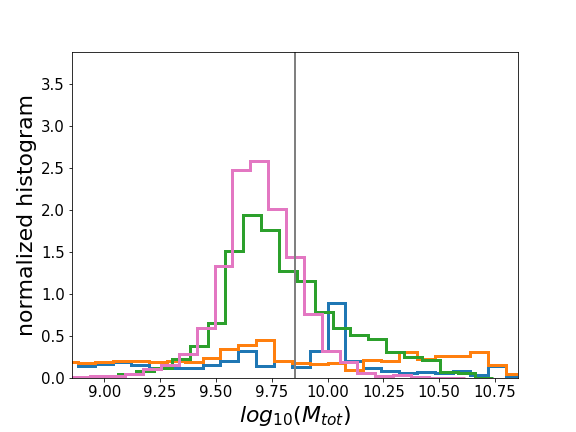,width=0.32\textwidth,clip=}
\epsfig{file=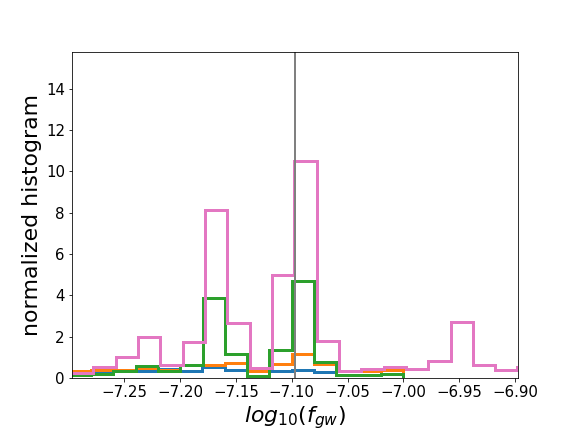,width=0.32\textwidth,clip=}
\epsfig{file=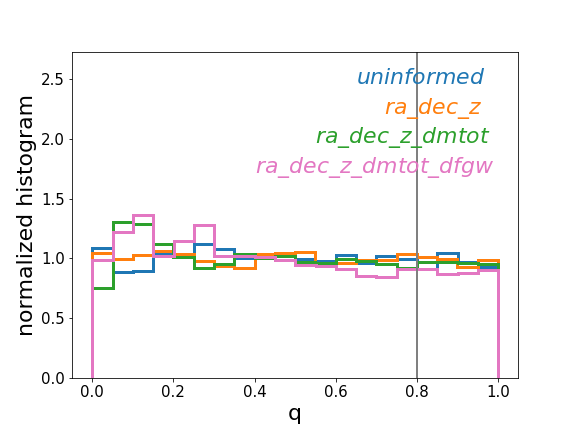,width=0.32\textwidth,clip=}

\epsfig{file=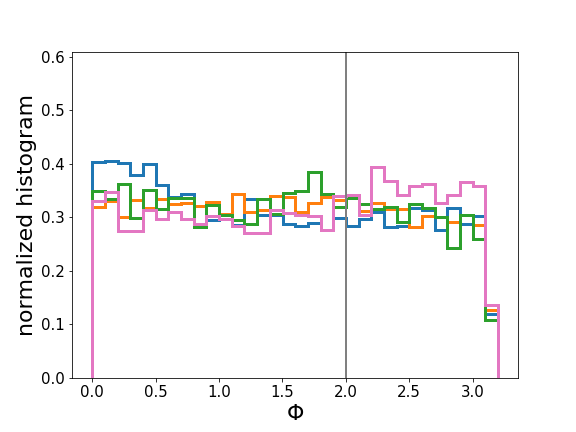,width=0.32\textwidth,clip=}
\epsfig{file=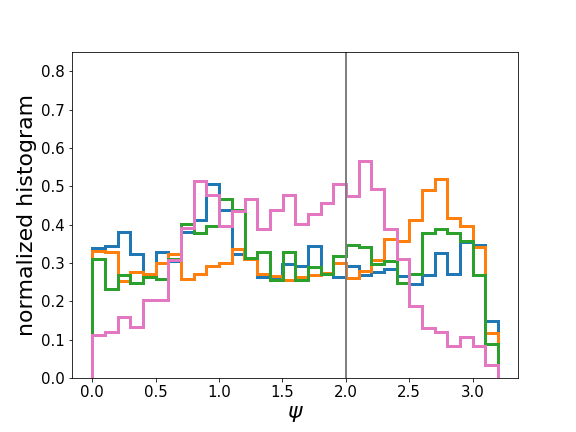,width=0.32\textwidth,clip=}
\epsfig{file=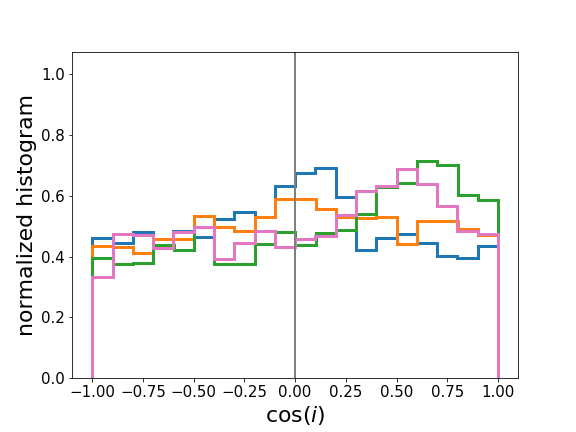,width=0.32\textwidth,clip=}

\caption{Same as Figure \ref{fig:tgt_sim91}, but for S-D2.}
\label{fig:tgt_sim98}
\end{figure*}

In this section, we focus on the ability of the PTA to estimate binary parameters, and the effects of an EM-informed targeted search on parameter estimation or limits at various sky locations and GW frequencies. We discuss the results separately for each injected signal (S-A to S-D), which are the posteriors we obtained for the parameters \{$M_{\rm tot}, q, f_{\rm gw}, i, \Phi_{0}, \psi$\} from the MCMC analysis. The posterior distributions are shown in Figures \ref{fig:tgt_sim31}-\ref{fig:tgt_sim98}. Note that the x-axis plotting ranges and histogram bin sizes for $M_{\rm tot}$ and $f_{\rm gw}$ vary between simulations S-A--S-D for presentation and are kept fixed between realizations (e.g., S-A1 and S-A2) for visual comparison.

To quantify the improvements in parameter estimation, we compute the width of the 90\% percent of the posterior distribution (90\% confidence interval, CI), which directly measures the precision of the parameter estimation. We also compute the Kullback-Leibler (KL) divergence \citep{Kullback1951}, which quantifies the difference between two probability distributions, and is given by

\begin{equation}
D_{\rm KL} = \sum_x{p(x) \ln \left[ \frac{p(x)}{q(x)} \right]} \, ,
\label{eqn:KL}
\end{equation}
    
\noindent where here $p(x)$ and $q(x)$ are the posterior and prior probability distributions of the parameter $x$. Therefore the KL divergence measures the difference between the prior used for a given parameter in a given search and the resultant posterior. A small value of $D_{\rm KL}$ indicates that the posterior is indistinguishable from the prior, whereas a larger value suggests that the parameter estimation is improved relative to the prior (recall that the parameters have different EM priors in the targeted searches). In Table \ref{tab:kl} and Table \ref{tab:kl2}, we show the 90\% CI and $D_{\rm KL}$ for each parameter from each search. A more detailed discussion of the KL divergence can be found in Appendix \ref{append:b}. We note that we calculate the KL divergence for each parameter in terms of the marginalized prior and posterior distributions, instead of a single N-dimensional KL divergence for all parameters. We also note that here we compute the KL divergence using the summation form (as opposed to integral form), because the posterior distributions cannot be fit to simple, known functions. Therefore, we choose the discrete form, to which the histograms of the priors and posteriors can be directly applied.

Simulation A (Figures \ref{fig:tgt_sim31} and \ref{fig:tgt_sim36}): S-A is located at a sensitive sky location and a sensitive GW frequency. In {\it uninformed}, evidence for a CW signal at the injected $f_{\rm gw}$ is very weak, with all parameters essentially unconstrained (blue histograms). Searching for the source at its injected sky location and distance ({\it ra\_dec\_z}) assists the algorithm in locating the correct $M_{\rm tot}$ (which manifests as a small peak in the posterior distribution, see the orange histogram in the top-left panel). It has also improved the estimation of $f_{\rm gw}$ (orange histogram in top center), even though both parameters are being searched over their respective full prior ranges. Applying prior information on $M_{\rm tot}$ in {\it ra\_dec\_z\_dmtot} slightly improves the actual estimation of the parameter (green histogram in top left), but improves the estimation of $f_{\rm gw}$ more significantly (green histogram in top center); see their respective $D_{\rm KL}$ values in Tables \ref{tab:kl} and \ref{tab:kl2}. The improvement on $f_{\rm gw}$ is worth noting, since the parameter is still being searched over the wide prior range at this stage. Including the additional prior on $f_{\rm gw}$ in {\it ra\_dec\_z\_dmtot\_dfgw} does not further improve the estimation of $M_{\rm tot}$ or $f_{\rm gw}$ significantly, as the green and purple histograms in the top-left and top-center panels, respectively, essentially converge and the $D_{\rm KL}$ value largely plateaus. Note that the estimation of $f_{\rm gw}$ greatly exceed the level of accuracy provided by the EM prior, which is indicated by the large final KL divergence (recall that the prior on $f_{\rm gw}$ is log-uniform with a factor of four uncertainty on $f_{\rm gw}$ on the linear scale). However, the mass ratio $q$ remains unconstrained in all searches (top right).

Additionally, we observe an overall improvement in the estimation of $\psi$ (bottom center), especially in {\it ra\_dec\_z\_dmtot} and {\it ra\_dec\_z\_dmtot\_dfgw} (green and purple histograms), even though we did not apply any priors on the parameter. It is therefore noteworthy that EM-informed priors on a subset of parameters have improved the estimation of the other parameters for which we have no EM information. However, $\Phi_{0}$ and $\cos i$ remain essentially unconstrained.

Simulation B (Figures \ref{fig:tgt_sim52} and \ref{fig:tgt_sim58}): S-B is located at the same sensitive sky location as S-A, but emitting at a higher GW frequency. All parameters are again unconstrained in {\it uninformed} (blue histograms). In the targeted searches, we observe a similar trend of enhanced parameter estimation for all parameters. Furthermore, for each parameter, the posterior distributions are in good agreement with each other and peak near the injected value (orange, green, and purple histograms), except for $q$, on which there are only lower limits. We note that these improvements are achieved despite that the source is quieter than S-A (in terms of the $\mathcal{F}_{p}$ statistic or SNR). A comparison of their KL divergence values shows that parameters in S-B are overall better constrained than S-A, especially the $f_{\rm gw}$ parameter, which has the highest KL divergence. This echos the better detectability of S-B compared to S-A (Section \ref{sec:detect}). Again this is likely due to the higher frequency of the S-B signal, which appears to be beneficial not only for detecting the signal, but the parameter estimation as well.

Simulation C (Figures \ref{fig:tgt_sim71} -- \ref{fig:tgt_sim77}) and Simulation D (Figures \ref{fig:tgt_sim91} -- \ref{fig:tgt_sim98}): S-C and S-D are the counterparts of S-A and S-B, respectively, at a less sensitive sky location. While the parameter estimation shares similar overall trends as S-A and S-B, we also observe some differences, which are mainly (1) the improvements are less pronounced overall and (2) S-C1 is the only simulation where the EM prior on $M_{\rm tot}$ allows the algorithm to place a lower limit on $q$ (green and purple histograms in the top-right panel of Figure \ref{fig:tgt_sim71}). 


\section{Implications for the multi-messenger studies of SMBHBs}\label{sec:discuss}
 
PTA searches for GWs from SMBHBs benefit considerably from EM priors, and the resulting posteriors are obtained in a fundamentally different way. This therefore has several interesting applications in the multi-messenger searches for SMBHBs (e.g., through confirming EM-selected SMBHB candidates) and the studies of these physical systems (e.g., through first estimating the source parameters). 


\pagebreak

\begin{figure*}
\centering

\epsfig{file=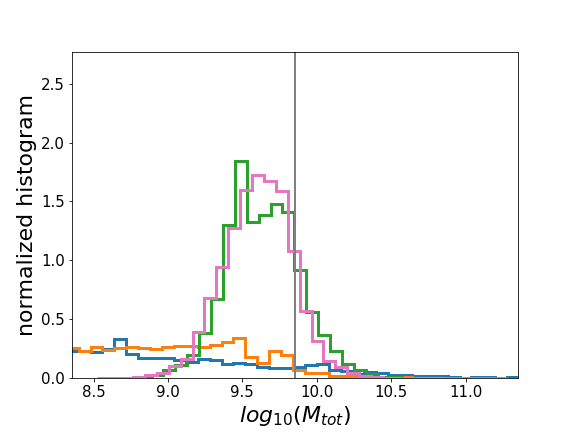,width=0.32\textwidth,clip=}
\epsfig{file=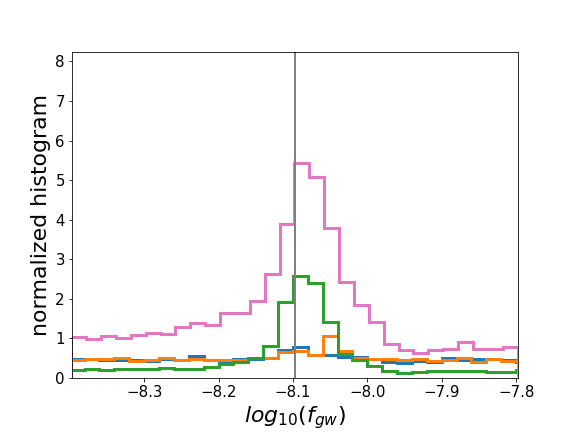,width=0.32\textwidth,clip=}
\epsfig{file=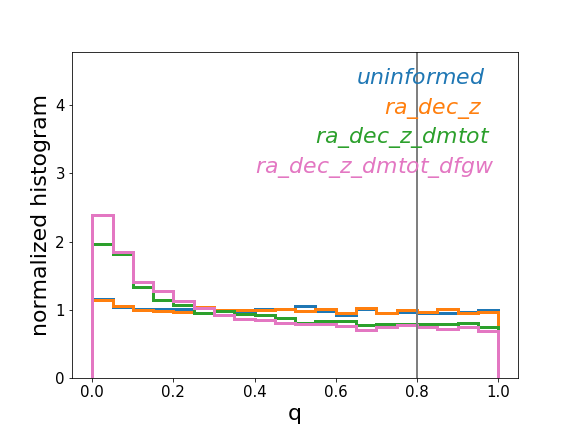,width=0.32\textwidth,clip=}

\epsfig{file=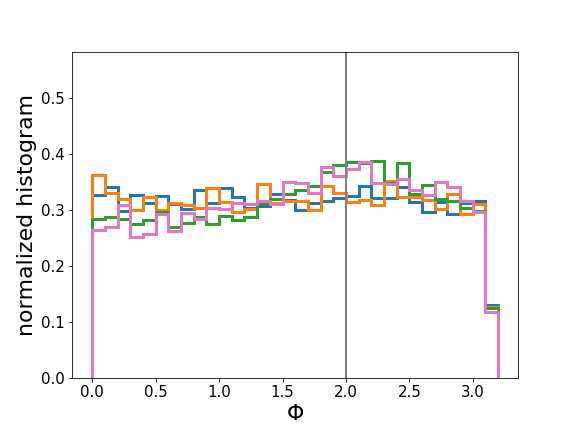,width=0.32\textwidth,clip=}
\epsfig{file=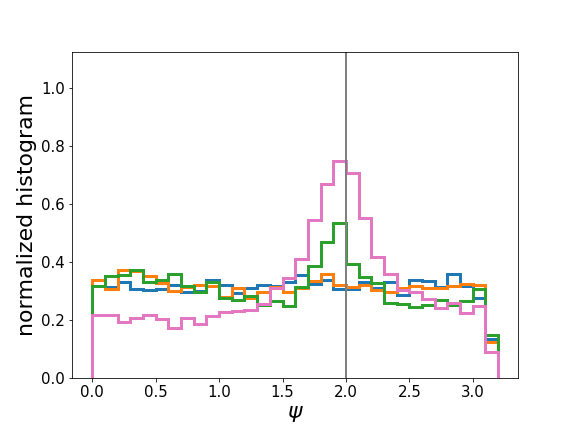,width=0.32\textwidth,clip=}
\epsfig{file=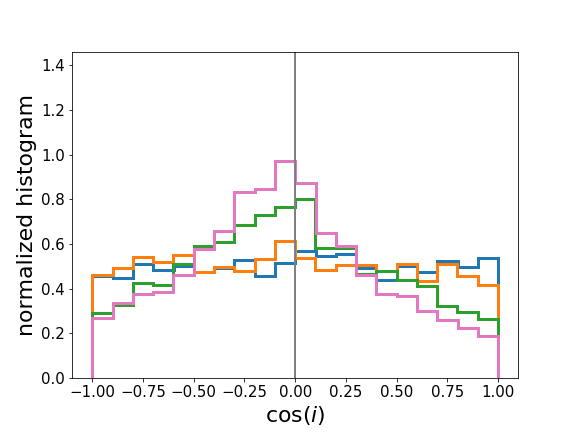,width=0.32\textwidth,clip=}

\caption{Same as Figure \ref{fig:tgt_sim31}, but for Simulation C1. The injected CW signal in S-C1 is located at a less favorable sky location and has a $f_{\rm gw}$ of 8 nHz.}
\label{fig:tgt_sim71}
\end{figure*}

\begin{figure*}
\centering

\epsfig{file=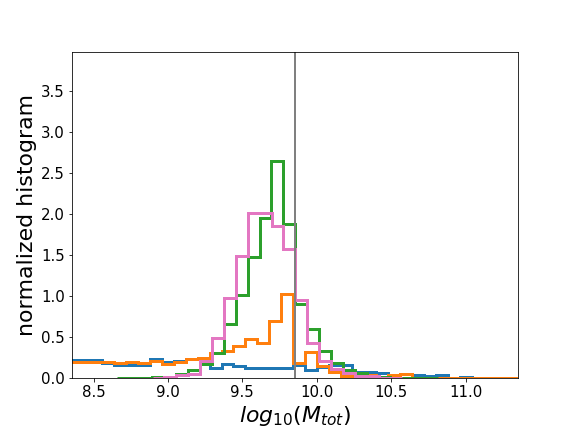,width=0.32\textwidth,clip=}
\epsfig{file=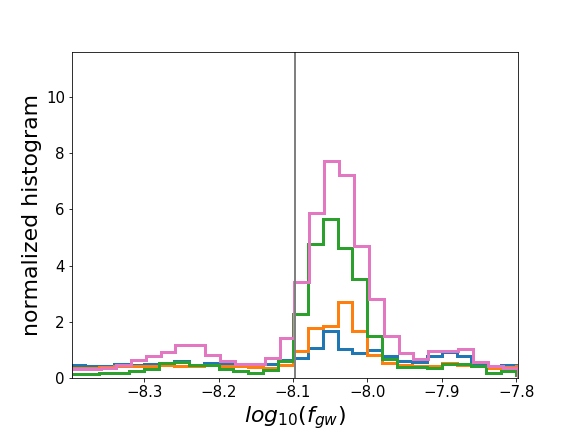,width=0.32\textwidth,clip=}
\epsfig{file=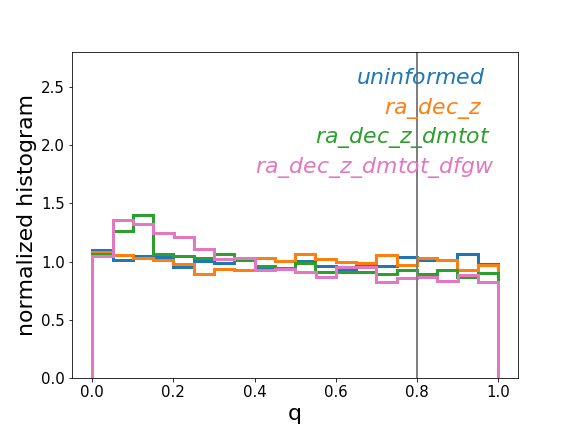,width=0.32\textwidth,clip=}

\epsfig{file=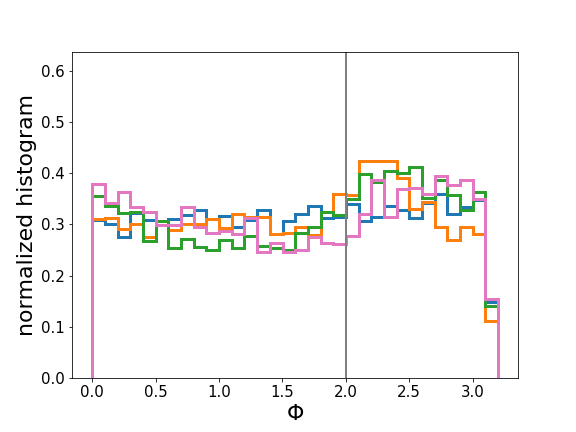,width=0.32\textwidth,clip=}
\epsfig{file=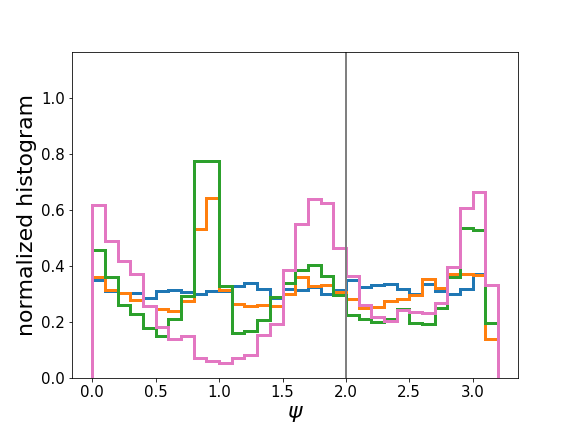,width=0.32\textwidth,clip=}
\epsfig{file=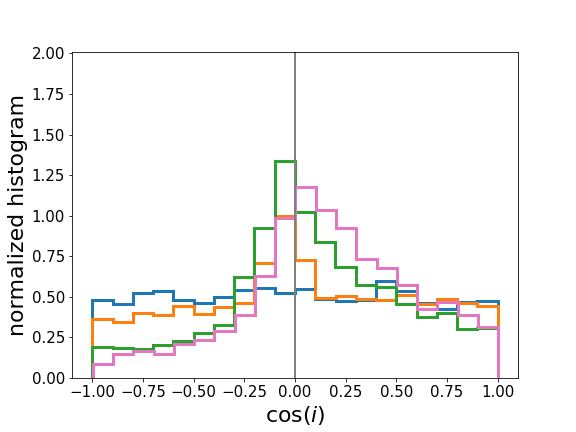,width=0.32\textwidth,clip=}

\caption{Same as Figure \ref{fig:tgt_sim71}, but for S-C2.}
\label{fig:tgt_sim77}
\end{figure*}

For a weak CW signal, which is likely the case for the first detection, knowing the source's location, redshift, and black hole mass through the observations of its EM counterpart significantly enhances the algorithm's ability to detect the source and estimate its parameters, which are otherwise unobtainable in an uninformed search. Therefore, using EM priors in GW searches could expedite the time-to-detection of individual sources. It remains to be seen, however, whether the same level of improvement can manifest for stronger GW signals.

Our analysis also suggests that the search for SMBHBs will benefit from this improvement regardless of their GW frequencies; there is also an indication that sources at higher frequencies may be more detectable and better characterized than those at lower frequencies. This will be a boon to the synergy and coordination between EM observers and the PTA community in several ways: (1) the time-domain search for GW-emitting SMBHBs typically requires at least a few cycles of quasi-periodicity in order to claim a possible binary origin and hence is more sensitive to SMBHBs with short periods (i.e., months--years timescales), or those that are emitting GWs at high frequencies (i.e., tens of nanohertz); (2) binary evolution and population estimates predict that long-period SMBHBs are more common (e.g., \citealt{Kelley2019, Krolik2019}), thus in general, the observational search for long-period SMBHBs is expected to be more fruitful. Since the improvement in parameter estimation is not sensitive to the GW frequency of the source, as suggested by our results, the multi-messenger search for SMBHBs is promising across the entire PTA band. 

By contrast, the improvement level does have a sky location dependence and is less pronounced if the source is located at a less favorable sky location, even though its nominal GW amplitude $h_{0}$ is larger. Therefore, the current NANOGrav PTA will have difficulty detecting and characterizing SMBHBs if they happen to be located where few pulsars are being monitored. Coincidentally, many of the current SMBHB candidates, including OJ 287, are located at such sky locations. Fortunately, as more pulsars across the sky are being added to the PTA, either as a regional PTA or under the umbrella of the International Pulsar Timing Array (IPTA; \citealt{IPTA}), this weakness will be overcome, PTA's synergies with EM facilities will strengthen, and there is great potential that it can operate as a truly all-sky nanohertz GW observatory.

The EM localization of an SMBHB (or candidate) usually implies that its black hole mass can also be measured. This additional EM prior, even with a considerable uncertainty, can enhance the CW search with the PTA even further. The most astrophysically interesting result is the ability to measure the GW frequency $f_{\rm gw}$ with PTA data at this stage, when it is otherwise completely unknown from EM observations alone. For a circular binary, which is assumed throughout the paper, this directly informs the orbital frequency $f_{\rm gw}$ = 2$f_{\rm orb}$. If a quasi-periodic variation of the AGN EM flux is observed in the SMBHB system, where the observed quasi-period (or frequency\footnote{The following discussion assumes $f_{\rm var}$ is in the rest frame.} $f_{\rm var}$) is interpreted as the electromagnetic imprint of the orbital period, then a direct comparison between $f_{\rm orb}$ and $f_{\rm var}$ is possible. For example, in the scenario that relativistic Doppler boost can be the cause of periodicity in an SMBHB \citep{D'Orazio2015}, which observationally favors an unequal mass ratio, the model has the clear prediction that $f_{\rm var}=f_{\rm orb}$. Through time-series analysis methods, $f_{\rm var}$ (or $f_{\rm orb, EM}$) can be determined to good accuracy, which can then be directly compared with $f_{\rm orb, GW}$. Through EM light-curve fitting, one additionally obtains an estimate of the mass ratio (albeit with large uncertainties), which can be compared with the lower limit on mass ratio that the PTA may already place (e.g., Figure \ref{fig:tgt_sim71} top-right panel) as additional evidence for the presence of a binary.

An alternative cause of periodicity in an SMBHB is the formation of a lump in the inner circumbinary disk which orbits at the local Keplerian rate $f_{\rm lump}$, interacts with passing BHs, and modulates the accretion of material at the beat frequency (e.g., \citealt{Farris2014, Noble2021}). Since this scenario can produce large-amplitude variations without strongly favoring low mass ratios, high black hole masses, or high inclinations, it is expected to be observationally more common \citep{Kelley2019}. However, translating from the observed $f_{\rm var}$ to the intrinsic $f_{\rm orb}$ is less straightforward than the previous scenario, as it depends on the binary mass ratio. Recent magneto-hydrodynamic simulations of varying mass ratios by \cite{Noble2021} confirm the previous, hydrodynamic-only results that only for $q\gtrsim0.2$ can the binary's gravitational torque be strong enough to excite a lump in the circumbinary disk which is at a radius of a few times the binary separation. For these intermediate mass ratios, the less massive, secondary BH predominantly interacts with the lump, therefore producing periodicity at $f_{\rm var}\approx f_{\rm orb}$. For an equal mass binary, the two BHs equally interact with the lump at the beat frequency $f_{\rm orb}-f_{\rm lump}$; the overall periodicity is thus at the frequency $f_{\rm var} = 2(f_{\rm orb}-f_{\rm lump})\approx 1.5f_{\rm orb}$. The EM-observed $f_{\rm var}$ and the independent GW measurement of the intrinsic $f_{\rm orb}$ (through $f_{\rm gw}$) would allow one to connect the two timescales, which informs a crude estimate of the mass ratio (i.e., equal mass ratio $q$ or small $q$) at no additional cost. Note that, again, this is independent from the mass ratio parameter estimation or limit obtained from the CW search. Even if it is a lower limit in the weak CW signal regime, it nevertheless has a considerable cross-validation value.

Finally, the overall ability of the PTA to cross-validate binary parameters, which is an integral part of confirming an EM-selected binary candidate, is excellent. Depending on the source's sky location and GW frequency, the PTA can estimate the parameters at a level which at least exceeds EM-alone (which may only be a factor of a few), and at the few percent level in some parameters. This suggests that even weak CW signals, which are the expected first signals and the majority of most CWs within the reach of a future PTA, have considerable multi-messenger science values. These are realistic but conservative assessment of the parameter estimation ability of PTAs, since the precision level for signals with larger SNRs is expected to be higher, as the SNR of the signal increases with a longer data span. Future work using projected PTA datasets can verify this predication.


\section{Summary and Conclusions} \label{sec:conclude}

In this work, we investigated the ability of the PTA to perform CW searches in the most realistic set-up yet, focusing on SMBHB parameter estimation in an uninformed search with GW data only and EM-informed searches in a multi-messenger scheme. To this end, we first constructed a simulated PTA dataset that resembles the 11 yr NANOGrav data set which takes into account both white noise and red noise of the pulsars as well as their non-uniform timing cadence and sky locations. We injected into the PTA dataset CW signals of modest amplitudes as proxies for the first CW signals in a future PTA dataset, which are likely to be weak and only marginally detectable. In contrast to previous work that studied CW detection or parameter estimation (e.g., \citealt{Sesana2010,Ellis2012}), the injected CW signal contains the full signal of the pulsar term, which allows us to include the astrophysically interesting chirp mass (or equivalently, total black hole mass) parameter in our analysis. We also vary the GW frequency and sky location of the CW signal, in order to study the effects they have on CW searches.

We then attempted to recover the injected signals in a procedure that mimics the search for a CW source in PTA data, using the Bayesian MCMC data analysis algorithm for CW detection. This approach thus gives a more realistic estimate for the precision and accuracy of binary parameter estimation than the lower limit estimates given by the Fisher information matrix approach \citep{Sesana2010}. Our main results can be summarized as the following:

\begin{enumerate}

\item In the weak signal regime, which is likely the case for the first CW signals emerging in PTA data, GW data alone struggle to recover the CW signal and its parameters.

\item Searching for the source using EM priors on its sky location, distance, and black hole mass can increase the Bayes factor by a factor of a few up to an order of magnitude if the source is located at a favorable sky location where more pulsars are being monitored. However, we do not observe the same effect if the source is located at a sky location with few pulsars. There are also hints that sources emitting at higher frequencies may be more detectable than those at lower frequencies.

\item By searching for the source at its sky location and distance through the observations of its EM counterpart, we are able to identify the correct parameter values, especially the astrophysically interesting $M_{\rm tot}$ and $f_{\rm gw}$, if the source is located at a favorable sky location. However, this improvement in binary parameter estimation does not manifest for a source located at a less favorable sky location. Additionally, similar to source detection, the parameter estimation of sources at higher GW frequencies also appears to be better.

\item Further including an EM prior on $M_{\rm tot}$ in the GW search results in the estimation of the parameter that is better than that from EM alone. Additionally, a prior on $M_{\rm tot}$ improves the constraints on the GW frequency, which is otherwise completely unknown from EM observations. Both effects are seen in Simulations A -- D, i.e., for both sensitive and less sensitive sky locations and frequencies, although the effects are stronger at the sensitive sky location.

\item If $f_{\rm gw}$ can be estimated within a factor of a few from EM observations, such as through analyzing the source's EM light curve, it can improve the estimation of all parameters even further. It may also permit the extraction of additional information about the system (such as an estimate of the mass ratio), if EM and GW data are interpreted jointly. This is also observed in Simulations A -- D, but is more pronounced at the sensitive sky location.

\item Overall, incorporating \emph{any} EM priors on a subset of binary parameters improves the estimation of all or almost all parameters, regardless of sky location and GW frequency.

\item Overall, combining EM and GW data results in parameter estimation at a precision level which is better than the two separately and at a level that permits the cross validation of binary parameters.

\end{enumerate}

These results highlight the complementarity of EM and GW data in searching for SMBHBs, and the feasibility of confirming an SMBHB candidate (by detecting the source and verifying its source parameters) and understanding the physical system in a multi-messenger approach. But here we have only scratched the surface of the potential of combining EM and GW data to advance our study of SMBHBs. Possible future work includes extending to higher SNR signals and applying the parameters of a more astrophysically motivated sample of SMBHBs. These inquiries will together form a more complete picture of the PTA as a GW detector and observatory of SMBHBs and give realistic estimates about its ability to study SMBHBs in a multi-messenger approach.


\acknowledgements
T.L. thanks members of the NANOGrav Detection and Astrophysics working groups, especially Bence B{\'e}csy, Steve Taylor, and Caitlin Witt, for helpful discussions. T.L. thanks Luke Kelley for a thorough reading of the manuscript and useful comments. T.L. acknowledges the use of the Geweke diagnostic which is implemented by Nihan Pol.

This work is supported by the NANOGrav National Science Foundation Physics Frontiers Center award no.~1430284. S.J.V. is supported by NSF award no.~2011772. This work was performed in part at the Aspen Center for Physics, which is supported by National Science Foundation grant PHY-1607611. The participation of T.L. at the Aspen Center for Physics was supported by the Simons Foundation.



\begin{table*}[ht]
\caption{90\% confidence interval and Kullback-Leibler divergence in uninformed and targeted searches (S-X1)}
\begin{center}
\begin{tabular}{ccccc}
\hline \hline
\multicolumn{5}{c}{Simulation A1 (90\%CI/D$_{\rm KL}$)}\\
\hline
 & {\it uninformed} & {\it ra\_dec\_z} & {\it ra\_dec\_z\_dmtot} & {\it ra\_dec\_z\_dmtot\_dfgw} \\
\hline
$\phi$ & 5.524/0.021 & \nodata & \nodata & \nodata \\
$\cos \theta$ & 1.801/0.014 & \nodata & \nodata & \nodata \\
$\log d_{\rm L}$ & 5.253/0.087 & \nodata & \nodata & \nodata \\
$\log M_{\rm tot}$ & 4.480/0.185 & 3.867/0.575 & 0.504/0.313 & 0.460/0.366 \\
$\log f_{\rm gw}$ & 1.776/0.135 & 1.666/0.616 & 0.574/2.162 & 0.103/1.555 \\
$q$ & 0.902/0.002 & 0.904/0.003 & 0.855/0.018 & 0.850/0.028 \\
$\Phi_{0}$ & 2.827/0.005 & 2.819/0.010 & 2.822/0.047 & 2.767/0.049 \\
$\psi$ & 2.847/0.004 & 2.741/0.046 & 2.460/0.279 & 2.171/0.343 \\
$\cos i$ & 1.794/0.011 & 1.817/0.006 & 1.769/0.014 & 1.785/0.017 \\
\hline

\multicolumn{5}{c}{Simulation B1 (90\%CI/D$_{\rm KL}$)}\\
\hline
 & {\it uninformed} & {\it ra\_dec\_z} & {\it ra\_dec\_z\_dmtot} & {\it ra\_dec\_z\_dmtot\_dfgw} \\
\hline
$\phi$ & 5.810/0.025 & \nodata & \nodata & \nodata \\
$\cos \theta$ & 1.825/0.015 & \nodata & \nodata & \nodata \\
$\log d_{\rm L}$ &  5.249/0.038 & \nodata & \nodata & \nodata \\
$\log M_{\rm tot}$ & 4.498/0.141  &  3.810/1.100 & 0.468/0.549 & 0.378/0.791 \\
$\log f_{\rm gw}$ & 1.822/0.175 & 1.796/1.352 & 1.644/2.741 & 0.014/3.188 \\
$q$ & 0.909/0.005 & 0.887/0.038  & 0.823/0.056 & 0.869/0.035 \\
$\Phi_{0}$ & 2.856/0.002 & 2.617/0.098 & 2.401/0.173 & 2.158/0.289 \\
$\psi$ & 2.803/0.005 & 2.607/0.180 & 1.903/0.477 & 1.168/0.754 \\
$\cos i$ & 1.801/0.005 & 1.579/0.126 & 1.267/0.253 & 0.980/0.456 \\
\hline

\multicolumn{5}{c}{Simulation C1 (90\%CI/D$_{\rm KL}$)}\\
\hline
 & {\it uninformed} & {\it ra\_dec\_z} & {\it ra\_dec\_z\_dmtot} & {\it ra\_dec\_z\_dmtot\_dfgw} \\
\hline
$\phi$ & 5.646/0.004 & \nodata & \nodata & \nodata \\
$\cos \theta$ & 1.791/0.003 & \nodata & \nodata & \nodata \\
$\log d_{\rm L}$ & 5.220/0.051 & \nodata & \nodata & \nodata \\
$\log M_{\rm tot}$ & 3.801/0.340 & 3.512/0.393 & 0.797/0.318 & 0.741/0.411 \\
$\log f_{\rm gw}$ & 1.804/0.012 & 1.796/0.015 & 1.794/0.428 & 0.486/0.225 \\
$q$ & 0.907/0.002 & 0.904/0.002 & 0.909/0.048 & 0.910/0.077 \\
$\Phi_{0}$ & 2.829/0.001 & 2.838/0.001 & 2.797/0.007 & 2.786/0.007 \\
$\psi$ & 2.812/0.003 & 2.833/0.003 & 2.829/0.020 & 2.700/0.107 \\
$\cos i$ & 1.795/0.005 & 1.774/0.005 & 1.652/0.056 & 1.597/0.106 \\
\hline

\multicolumn{5}{c}{Simulation D1 (90\%CI/D$_{\rm KL}$)}\\
\hline
 & {\it uninformed} & {\it ra\_dec\_z} & {\it ra\_dec\_z\_dmtot} & {\it ra\_dec\_z\_dmtot\_dfgw} \\
\hline
$\phi$ & 5.821/0.014 & \nodata & \nodata & \nodata \\
$\cos \theta$ & 1.800/0.008 & \nodata & \nodata & \nodata \\
$\log d_{\rm L}$ & 5.384/0.042 & \nodata & \nodata & \nodata \\
$\log M_{\rm tot}$ & 4.080/0.267 & 3.742/0.301 & 0.841/0.120 & 0.609/0.317 \\
$\log f_{\rm gw}$ & 1.835/0.047 & 1.815/0.030 & 1.824/0.355 & 0.462/0.765 \\
$q$ & 0.910/0.002 & 0.905/0.001 & 0.905/0.011 & 0.900/0.006 \\
$\Phi_{0}$ & 2.821/0.001 & 2.817/0.001 & 2.841/0.001 & 2.806/0.009 \\
$\psi$ & 2.815/0.004 & 2.820/0.002 & 2.802/0.011 & 2.709/0.097 \\
$\cos i$ & 1.792/0.004 & 1.784/0.003 & 1.753/0.015 & 1.624/0.121 \\

\hline \hline
\end{tabular}
\end{center}
\label{tab:kl}
\end{table*}

\begin{table*}[ht]
\caption{Same as Table \ref{tab:kl}, but for S-X2}
\begin{center}
\begin{tabular}{ccccc}
\hline \hline
\multicolumn{5}{c}{Simulation A2 (90\%CI/D$_{\rm KL}$)}\\
\hline
 & {\it uninformed} & {\it ra\_dec\_z} & {\it ra\_dec\_z\_dmtot} & {\it ra\_dec\_z\_dmtot\_dfgw} \\
\hline
$\phi$ & 5.559/0.016 & \nodata & \nodata & \nodata \\
$\cos \theta$ & 1.793/0.006 & \nodata & \nodata & \nodata \\
$\log d_{\rm L}$ & 5.287/0.048 & \nodata & \nodata & \nodata \\
$\log M_{\rm tot}$ & 4.229/0.227 & 3.557/1.031 & 0.565/0.262 & 0.541/0.320 \\
$\log f_{\rm gw}$ & 1.780/0.047 & 1.348/0.726 & 0.581/1.656& 0.290/1.256 \\
$q$ & 0.903/0.002 & 0.878/0.007 & 0.866/0.022& 0.886/0.011\\
$\Phi_{0}$ & 2.838/0.001 & 2.767/0.009 & 2.824/0.034 &2.779/0.024 \\
$\psi$ & 2.850/0.006 & 2.786/0.039 & 2.605/0.140 & 1.927/0.347 \\
$\cos i$ & 1.808/0.004 & 1.807/0.013 & 1.790/0.010 &1.762/0.013  \\
\hline

\multicolumn{5}{c}{Simulation B2 (90\%CI/D$_{\rm KL}$)}\\
\hline
 & {\it uninformed} & {\it ra\_dec\_z} & {\it ra\_dec\_z\_dmtot} & {\it ra\_dec\_z\_dmtot\_dfgw} \\
\hline
$\phi$ & 5.563/0.005  & \nodata & \nodata & \nodata \\
$\cos \theta$ & 1.797/0.004 & \nodata & \nodata & \nodata \\
$\log d_{\rm L}$ & 5.289/0.044 & \nodata & \nodata & \nodata \\
$\log M_{\rm tot}$ & 4.033/0.312 & 0.330/2.911 & 0.193/1.399 & 0.264/1.129 \\
$\log f_{\rm gw}$ & 1.817/0.023 & 0.010/3.894 & 0.010/3.910 & 0.010/3.915 \\
$q$ & 0.906/0.003 & 0.860/0.086 & 0.776/0.116 & 0.826/0.061 \\
$\Phi_{0}$ & 2.817/0.001 & 1.996/0.332 & 2.142/0.318 & 1.901/0.368 \\
$\psi$ & 2.826/0.003 & 0.978/1.007 & 1.115/0.927 & 0.968/0.992 \\
$\cos i$ & 1.797/0.005 & 1.175/0.368 & 1.189/0.354 &  1.173/0.378 \\
\hline

\multicolumn{5}{c}{Simulation C2 (90\%CI/D$_{\rm KL}$)}\\
\hline
 & {\it uninformed} & {\it ra\_dec\_z} & {\it ra\_dec\_z\_dmtot} & {\it ra\_dec\_z\_dmtot\_dfgw} \\
\hline
$\phi$ & 5.591/0.012 & \nodata & \nodata & \nodata \\
$\cos \theta$ & 1.788/0.005 & \nodata & \nodata & \nodata \\
$\log d_{\rm L}$ & 5.173/0.061 & \nodata & \nodata & \nodata \\
$\log M_{\rm tot}$ &  3.955/0.295& 3.686/0.436 & 0.661/0.288 & 0.619/0.376 \\
$\log f_{\rm gw}$ &  1.754/0.061 & 1.737/0.189 &1.275/1.029  & 0.406/0.533 \\
$q$ & 0.905/0.002 &  0.903/0.004& 0.897/0.009 & 0.892/0.015 \\
$\Phi_{0}$ & 2.835/0.003 & 2.794/0.010 & 2.859/0.014  & 2.862/0.012 \\
$\psi$ &2.854/0.004  & 2.860/0.034&  2.929/0.110& 3.001/0.200 \\
$\cos i$ & 1.789/0.008 &1.744/0.039  &1.570/0.184  & 1.478/0.209 \\
\hline

\multicolumn{5}{c}{Simulation D2 (90\%CI/D$_{\rm KL}$)}\\
\hline
 & {\it uninformed} & {\it ra\_dec\_z} & {\it ra\_dec\_z\_dmtot} & {\it ra\_dec\_z\_dmtot\_dfgw} \\
\hline
$\phi$ & 5.797/0.038 & \nodata & \nodata & \nodata \\
$\cos \theta$ &1.822/0.033  & \nodata & \nodata & \nodata \\
$\log d_{\rm L}$ &5.057/0.110  & \nodata & \nodata & \nodata \\
$\log M_{\rm tot}$ &4.103/0.282  &4.406/0.204  &0.927/0.085  &0.567/0.379  \\
$\log f_{\rm gw}$ &1.796/0.147  &1.853/0.069  &1.822/0.377  &0.494/0.674  \\
$q$ &0.902/0.003  &0.901/0.002  &  0.888/0.008 &0.893/0.011  \\
$\Phi_{0}$ &2.849/0.008  &2.812/0.002  &2.818/0.005  &2.853/0.006  \\
$\psi$ &2.856/0.018  &2.837/0.022  &2.780/0.020 &2.256/0.142  \\
$\cos i$ &1.781/0.021  &1.782/0.007  &1.781/0.029  & 1.754/0.017  \\
\hline

\hline \hline
\end{tabular}
\end{center}
\label{tab:kl2}
\end{table*}

\clearpage

\appendix
\restartappendixnumbering
\section{Bayes Factors of Uninformed and Targeted Searches (S-A -- S-D)}\label{append:a}

We computed the Bayes factors shown in Fig.~\ref{fig:sim_bf} using the Savage-Dickey (SD) formula \citep{Dickey1971}, which can be used to compute the Bayes factor for two nested models. In this case, the two models are a CW signal-plus-noise model $\mathcal{H}_1$ and a noise-only model $\mathcal{H}_0$, which is equivalent to model $\mathcal{H}_1$ with $h_{0}=0$. The Bayes factor $\mathcal{B}_{10}$ is given by

\begin{equation}
    \mathcal{B}_{10} \equiv \frac{\mathrm{evidence}[\mathcal{H}_1]}{\mathrm{evidence}[\mathcal{H}_0]} = \frac{p(h_{0}=0|\mathcal{H}_1)}{p(h_{0}=0|\mathcal{D}, \mathcal{H}_1)} \,,
\end{equation}

\noindent where $p(h_{0}=0|\mathcal{H}_1)$ is the prior probability density of $h_{0} = 0$ in the embedding model $\mathcal{H}_1$, and $p(h_{0}=0|\mathcal{D}, \mathcal{H}_1)$ is the posterior probability density of $h_{0} = 0$ in the embedding model $\mathcal{H}_1$.

In order to use the SD formula, we first computed the prior and posterior distributions of $h_{0}$ using the distributions of $M_{\rm tot}$, $q$, $f_{\rm gw}, and d_{\rm L}$ for each mock search in each simulated dataset, using the relationship

\begin{equation}
	\log_{10}\mathcal{M} = \frac{3}{5} \log_{10}\left[\frac{q}{(1+q)^2}\right] + \log_{10} M_\mathrm{tot} \,,
\end{equation}
\begin{equation}
	h_{0} = \frac{2\mathcal{M}^{5/3}(\pi f_{\rm gw})^{2/3}}{d_{\rm L}} \,.
\end{equation}

\noindent The Bayes factor is then computed by taking the ratio of the numbers of samples below some threshold between the prior and the posterior.

\begin{figure*}[hb]
\centering
\epsfig{file=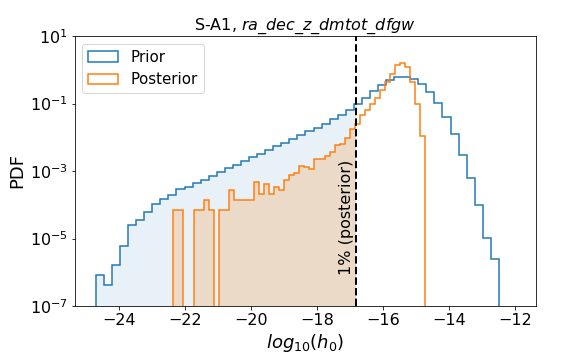,width=0.5\textwidth,clip=}
\caption{The prior and posterior distributions of $h_{0}$ for Simulation A1, {\it ra\_dec\_z\_dmtot\_dfgw} (blue and orange histograms, respectively). The black dashed line marks the one percentile of samples in the posterior, which is used as the threshold to compute the SD ratio, which is the ratio between the blue- and orange-shaded areas.}
\label{fig:sa_h0}
\end{figure*}

We adopt the threshold in terms of the percentile of samples in the posterior. We then use the $h_{0}$ value corresponding to this percentile to determine the number of samples below this threshold in the prior. The reason is two-fold: first, it guarantees a sufficient number of samples below the threshold and is more robust agains imperfect sampling, and second, it is calculated in a self-consistent manner for all simulations and realizations, while still preserving the principle behind the SD approximation. We then repeat the procedure by varying the threshold between 1\% and 5\%, both of which amplitudes are sufficiently low for the SD approximation to be valid. The uncertainty of $\mathcal{B}_{10}$ is computed from the variance in $\mathcal{B}_{10}$ as the threshold is varied. For illustration purposes, Figure~\ref{fig:sa_h0} shows the prior and posterior distributions for $h_{0}$ for S-A1, {\it ra\_dec\_z\_dmtot\_dfgw}. The SD ratio is then the ratio between the blue and orange-shaded areas.

\clearpage
\section{KL Divergence} \label{append:b}

The KL divergence measures the difference between two probability distributions:

\begin{equation}
    D_\mathrm{KL}(P||Q) = \sum_x P(x) \log\left[\frac{P(x)}{Q(x)}\right] \,,
\end{equation}

\noindent where $P(x)$ and $Q(x)$ are discrete probability distributions. In Sec.~\ref{sec:results}, we take $P(x)$ to be the posterior probability density function, and $Q(x)$ to be the prior probability density function. Then the KL divergence can be interpreted as the information gained from the posterior compared to the prior.

In order to aid in understanding the significance of the values of the KL divergence in Table~\ref{tab:kl} and Table \ref{tab:kl2}, here we work through a toy example. Suppose we are performing parameter estimation for a one-parameter model and obtain a Gaussian posterior

\begin{equation}
    p(x) = \frac{1}{\sqrt{2\pi\sigma}} \exp \left[-\frac{(x-\mu)^2}{2\sigma^2}\right] \,,
\end{equation}

\noindent where $\mu$ is the mean and $\sigma$ is the variance. Assume the prior for this parameter is uniform over the range $[\mu - a, \mu + a]$:

\begin{equation}
    q(x) = \left\{ \begin{array}{cc} \frac{1}{2a} & \mu-a \leq x \leq \mu+a \\ 0 & \mathrm{otherwise} \end{array} \right.
\end{equation}

\noindent For continuous probability distributions, the KL divergence is given by 

\begin{equation}
    D_\mathrm{KL}(P||Q) = \int_{-\infty}^\infty dx \; p(x) \log\left[\frac{p(x)}{q(x)}\right] \,.
\end{equation}

\noindent For the posterior and prior given above, the KL divergence can be computed analytically:

\begin{equation}
    D_\mathrm{KL}(P||Q) = \frac{a}{\sqrt{2\pi\sigma^2}} \exp\left(-\frac{a^2}{2\sigma^2}\right) + \frac{1}{2} \mathrm{erf}\left(\frac{a}{\sqrt{2\sigma^2}}\right) \left[ \log\left(\frac{2a^2}{\pi\sigma^2}\right) - 1 \right] \,.
\end{equation}

\begin{figure}[hb]
\centering
\epsfig{file=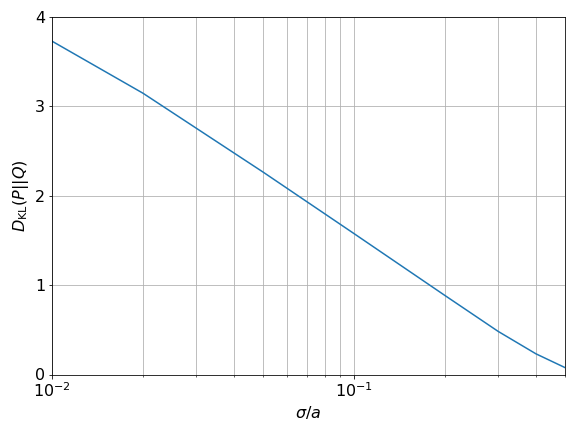,width=0.45\textwidth,clip=}
\caption{KL divergence $D_\mathrm{KL}(P||Q)$ for a posterior $p$ and uniform prior $q$. The posterior is assumed to be Gaussian 
with mean $\mu$ and variance $\sigma$, while the prior is assumed to be uniform centered around $\mu$ with width $2a$. The KL divergence measures the additional information in the posterior compared to the prior: larger values correspond to a more constrained posterior.}
\label{fig:KL_toy_problem}
\end{figure}

\noindent Figure~\ref{fig:KL_toy_problem} shows the KL divergence as a function of the variance of the Gaussian posterior $\sigma$.

We emphasize that this is a toy problem, and the posteriors shown in Sec.~\ref{sec:results} are not simple Gaussian distributions. However, these values may help the reader understand the significance of the values in Table~\ref{tab:kl} and Table \ref{tab:kl2}.

\clearpage

\end{document}